\documentclass[fleqn,usenatbib]{mnras}
\usepackage{newtxtext,newtxmath}
\usepackage[T1]{fontenc}
\usepackage{ae,aecompl}

\usepackage{amsmath}	

\usepackage[dvipsnames]{xcolor}

\usepackage{graphicx}
\usepackage{subfigure}
\newcommand\msun{{\,M_\odot}}

\newcommand\zsun{{\rm \,Z_\odot}}

\usepackage{aecompl}
\usepackage[T1]{fontenc}
\usepackage{threeparttable}
\usepackage{caption}
\usepackage{times}
\usepackage{ifpdf}
\usepackage{ulem}

\usepackage{float}

\newcommand{\cmci}{~\mbox{cm}^{-3}}

\newcommand{\Msun}{~\mbox{M}_{\odot}}

\title[Hypernova signatures of the first stars in dwarfs]{Hypernova signatures of the first stars in dwarf galaxies in the Local Group}

\author[T. Y. Lee et al.]{Teayong Lee$^{1}$, Myoungwon Jeon$^{1,2}$\thanks{E-mail:myjeon@khu.ac.kr} and  
Volker Bromm$^{3,4}$\\
$^{1}$School of Space Research, Kyung Hee University, 1732 Deogyeong-daero, Yongin-si, Gyeonggi-do 17104, Korea\\
$^{2}$Department of Astronomy \& Space Science, Kyung Hee University, 1732 Deogyeong-daero, Yongin-si, Gyeonggi-do 17104, Korea\\
$^{3}$Department of Astronomy, University of Texas, Austin, TX 78712, USA\\
$^{4}$Weinberg Institute for Theoretical Physics, University of Texas at Austin, Austin, TX 78712, USA
}

\date{Accepted XXX. Received YYY; in original form ZZZ}

\pubyear{2021}

\begin{document}
\label{firstpage}
\pagerange{\pageref{firstpage}--\pageref{lastpage}}
\maketitle

\begin{abstract}
Observing the first generation of stars, Population III (Pop III), is still a challenge even with the James Webb Space Telescope (JWST) due to their faintness. Instead, searching for fossil records of Pop III stars in nearby dwarf galaxies provides an alternative method for studying their physical properties. It is intriguing that a star recently discovered in the Sculptor dwarf galaxy, named AS0039, is considered to show the unique signature of a Pop~III star. The detailed abundance patterns of AS0039 are well-matched with those predicted by nucleosynthesis models for Pop~III exploding as an energetic hypernova (HN), confirming its potential to provide insight into the properties of the first stars. This study aims to explore the environmental conditions required for the formation of such a unique star using cosmological hydrodynamic zoom-in simulations on dwarf galaxies with a mass of $M_{\rm vir}\approx10^8\msun$ at $z=0$ while varying the fraction of Pop~III stars that undergo HNe. Our simulations identify rapid gas inflow ($\dot{M}_{\rm gas}\sim0.08\msun$ $\rm yr^{-1}$) as a possible factor in facilitating the formation of stars similar to AS0039. Alternatively, the delayed formation of subsequent Pop~II stars in the gas-enriched environment may lead to low-metallicity stars like AS0039. Additionally, using the {\sc A-SLOTH} code, we investigate the probability of finding remnants of Pop II stars with HN signatures in nearby dwarf satellite galaxies. We suggest that the most likely dwarf galaxies to contain HN signatures are massive satellites with a probability of 40\% in the range of $M_{\rm peak}\approx10^{10}-10^{11}\msun$ and $M_{\ast}\approx10^7-10^8\msun$, considering observational limitations.
\end{abstract}




\section{Introduction}

Understanding the physical characteristics of the first generation of stars, known as Population III (Pop III) stars, and the first galaxies is crucial for gaining a comprehensive understanding of the entire Universe (reviewed in \citealp{Bromm2009, Bromm2013, Klessen2023}). Recent breakthroughs in observing distant galaxies born at $z\gtrsim10$ by the James Webb Space Telescope (JWST) have opened up new previously inaccessible territories of the early Universe (e.g., \citealp{Atek2022, Donnan2022, Whitler2022, Bouwens2022, Finkelstein2022}). The discovery of unexpectedly bright high-$z$ galaxies has prompted the investigation of the possible existence of massive Pop~III stars in these galaxies (e.g., \citealp{Kannan2023, Boylan-Kolchin2022, Inayoshi2022, Haslbauer2022}). As a result, understanding the characteristics of the first stars has become even more important. While JWST could possibly directly observe the first stars, if Pop~III were making up most of the stellar mass in early galaxies or globular clusters (e.g., \citealp{Mowla2022}), detecting individual Pop~III stars remains challenging and requires larger telescopes (e.g., \citealp{Schauer2020, Woods2021, Katz2023, Larkin2023, Venditti2023}), except for the cases where the brightness is significantly increased by gravitational lensing effects (e.g., \citealp{Schauer2022, Welch2022}).

\par
Although direct detection of the first stars remains elusive, extensive theoretical studies have been conducted to infer their physical properties (see, e.g., \citealp{Bromm2013} for a review). It is widely accepted that Pop III stars were massive stars, with masses larger than a few tens of solar masses, formed from primordial gas in minihaloes with virial masses of $M_{\rm vir}\approx10^{5-6}\msun$ at $z\gtrsim15$ 
 (e.g., \citealp{Haiman1997, Tegmark1997, Bromm1999, Abel2002, Yoshida2003}). Pop~III stars, depending primarily on their initial masses, are expected to undergo supernovae (SNe) explosions, ejecting metals synthesized in their cores and polluting the surrounding interstellar medium (ISM) (e.g., \citealp{Wise2008, Greif2010, Wise2012, Jeon2014}). This contaminated medium eventually gives rise to the second generation of stars, Population~II (Pop~II) stars, which tend to have lower masses and longer lifespans than their Pop~III predecessors (e.g., \citealp{Omukai2000, Bromm2001b, Schneider2002}). Consequently, such Pop~II stars ($m_{\rm PopII}\lesssim0.8\msun$) may still exist in the Local Universe, displaying the distinctive characteristics of Pop~III stars. The search for such traces of ancient stars that existed in the Local Universe is known as ``stellar archaeology." Furthermore, if these Pop~II stars belong to local dwarf galaxies, it is possible to infer information not only about the stars themselves but also about the environment in which they formed by reconstructing the star formation histories (SFHs) of these galaxies. This approach is known as ``galaxy archaeology" (e.g., \citealp{Bovill2011, Frebel2015})

\par

\par
What unique characteristics of the first generation of stars have been preserved in local dwarf galaxies? One possible answer lies in carbon-enhanced metal-poor stars (CEMPs) (e.g., \citealp{Beers2005, Aoki2007}), which have been thought to be associated with Pop~III stars due to their commonness in observed dwarf galaxies such that the fraction of CEMPs increases as decreasing metallicities, reaching $\sim$92\% below $\rm [Fe/H]<-4$ (e.g., \citealp{Lee2013, Placco2014, Placco2021}). Specifically, these stars are defined by $\rm [C/Fe]\gtrsim$ 0.7 and $\rm [Fe/H]\lesssim-2$. One possible explanation for this is the weak supernova explosion of Pop III stars, which can expel light elements such as carbon and oxygen from the outer layers of the star while heavy elements like iron fall back onto the core (e.g., \citealp{Norris2013, Jeon2021a}).

\par
Recently, in the Sculptor dwarf spheroidal (dSph), \citet{Skuladottir2021} have identified a potential remnant of Pop~III stars by discovering AS0039, the most metal-poor star with $\rm [Fe/H]=-4.11$, among the observed local dwarf galaxies. According to the nucleosynthesis model for metal-free stars proposed by \citet{Heger2002, Heger2010}, the chemical abundance patterns of AS0039 appear to be consistent with those of a Pop~III star with a mass of $m_{\rm PopIII}\approx20\pm2\msun$ that ended its life as a hypernova (HN) explosion with an SN energy of $E_{\rm SN}=10^{52}$ erg. Notably, {\sc AS0039} is a peculiar carbon-poor star with $\rm [C/Fe]_{LTE}\sim -0.75$ and has distinct $\alpha$-element patterns, which sets it apart from normal-carbon stars. According to \citet{Hartwig2023}, there is a 77\% chance that the enrichment of {\sc AS0039} is due to a single Pop~III SN event. Furthermore, \citet{Placco2021} have reported another potential HN signature associated with Pop III stars. They have identified a star in Stripe 82 called {\sc SPLUS J2104-0049}, whose chemical composition appears to be consistent with that of a Pop~II star formed from gas polluted by a Pop III star with a mass of $30\msun$. This Pop III star would have exploded as an HN with an energy of $E_{\rm SN}=10^{52}$ erg as well. 

\par
The aim of this study is to offer a theoretical interpretation of the observed stars and gain insights into the era of the first and second generations of stars. In particular, the goal is to reproduce stars such as AS0039, which are rare and could potentially offer clues about  Pop~III HN events as a component of the stellar population in a nearby dwarf galaxy. To be specific, we investigate how the signatures of Pop~III HN events might appear in subsequent Pop~II stars. Moreover, we examine the circumstances under which a star like the observed AS0039 can form. To achieve this goal, we conduct a series of cosmological zoom-in simulations on a local dwarf analog, which is comparable to the ultra-faint dwarf galaxy (UFD) with a mass of $M_{\rm vir}\lesssim10^8\Msun$ at $z=0$. UFDs are the smallest and most metal-poor galaxies in the Universe, and they are considered an excellent laboratory for studying the earliest stars since they preserve traces of them. (For a review, see \citealp{Simon2019}, and also see \citealp{Tolstoy2009, Brown2014, McConnachie2012}.)

\par
The exact mechanism for the explosion of highly energetic HN is still unclear. However, there is widespread agreement that it may be connected to massive stars exhibiting high angular momentum (e.g., \citealp{Nomoto2006, Burrows2007}). Several studies have been conducted on the level of rotation in metal-free stars (e.g., \citealp{Stacy2011, Hirano2018}). One such study by \citet{Stacy2011} estimates the rotational velocity of Pop~III stars by tracking the evolution of primordial gas up to densities of $n_{\rm H}=10^{12}\rm cm^{-3}$ in minihaloes. This suggests that the rotational velocity of stars larger than $30\msun$ may potentially exceed 1000 km $\rm s^{-1}$. Consequently, these stars may experience chemically homogeneous evolution (CHE) (\citealp{Sibony2022}), creating a reservoir of rotational energy sufficient to initiate an HN explosion. Another study by \citet{Choplin2019} investigated the abundance patterns of carbon-enhanced extremely metal-poor (EMP) stars ($\rm -4<[Fe/H]<-3$), comparing them to massive stellar models that consider rotation. The findings suggest a higher fraction of fast rotators at low metallicity, with the velocity distribution of star models reaching equatorial velocities of 550-640 km $\rm s^{-1}$. As such, whether or not Pop~III stars undergo HN explosions are determined by their initial masses as well as by their degree of rotation (e.g., \citealp{Marigo2003, Ekstrom2008, Heger2010, Chatzopoulos2012, Choplin2019, Murphy2021}). For instance, \citet{Yoon2012}, where they take into account magnetic fields and the rotation of metal-free stars, suggest that zero-age main sequence stars with masses between $13\msun$ and $84\msun$ may undergo HN explosions if they experience CHE. 

In this work, we have conducted cosmological simulations on a dwarf galaxy analog, where we vary the fraction of Pop~III stars that end their lives as HNe as a free parameter, covering the full range of possibilities. To further understand the observed frequency of HN signatures, we have also utilized a semi-analytic model of galaxy formation called {\sc A-SLOTH} (\citealp{Hartwig2022}). This model enables us to connect the formation of early Pop~III stars with their fossil remnants in nearby galaxy analogs, providing an estimate of the probability of discovering satellite galaxies that contain Pop~III HN signatures based on the fraction of Pop~III stars exploding as HNe in the early Universe.

The paper is structured as follows: Section 2 outlines the numerical methodology employed in this research. Section 3 presents and analyzes the simulation results, including the growth of simulated galaxies over cosmic time, the prevalence of minihaloes and Pop~II stars exhibiting Pop~III HN signatures, and the chemical imprints left on the remaining Pop~II stars. In Section 4, we discuss the likelihood of finding satellite galaxies that harbor Pop~III HN signatures using a semi-analytic galaxy formation model. The key discoveries of this study are summarized in Section 5. Unless otherwise indicated, all distances are provided in physical (proper) units to maintain consistency.

\section{Numerical methodology}
\label{Sec:Metho}

\subsection{Simulation Setup}
We have conducted cosmological hydrodynamic zoom-in simulations on dwarf galaxies using a modified version of the parallel N-body/smoothed particle hydrodynamics code, GADGET3 (e.g., \citealp{Springel2001, Springel2005, Schaye2010}). We utilize cosmological parameters with a fraction of dark energy of $\Omega_{\rm \Lambda}=0.73$, dark matter of $\Omega_{\rm m}=0.26$, baryons of $\Omega_{\rm b}=0.04$, and a Hubble constant of $\rm H_0=0.71$ km $\rm s^{-1}Mpc^{-1}$, respectively (\citealp{planck2016}). We employ the code MUSIC (\citealp{Hahn2011}) to generate initial conditions for a cosmological simulation box with a linear size of 3.125 comoving Mpc. In the first step, we conducted a dark-matter-only simulation to $z=0$ using $128^3$ particles to represent this box and then identify a target galaxy with a mass of $M_{\rm vir}\approx10^8\msun$ at $z=0$ using a halo-finder code called {\sc ROCKSTAR} (\citealp{Behroozi2013}). As a second step, we performed three consecutive refinements for the Lagrangian volume that encompasses particles within $R=2.5 R_{\rm vir}$ at $z=0$. This results in the most refined region where the masses of dark matter and gas particles are $m_{\rm DM}\sim 500\msun$ and $m_{\rm gas}\sim63\msun$, respectively. 

At each time step, we solve the rate equations for primordial species such as $\rm H, H^{+}, H^{-}, H_{2}, H^{+}2, He, He^{+}, He^{++}, e^{-}, D, D^{+}$, and HD, considering all the relevant cooling processes such as H and He collisional ionization, excitation and recombination cooling, bremsstrahlung, inverse Compton cooling, and collisional excitation cooling of $\rm H{2}$ and HD. We also account for the cooling of gas by metal species such as carbon, oxygen, silicon, magnesium, neon, nitrogen, and iron, using cooling rates from the photoionization package CLOUDY \citep{Ferland1998}. To mimic the effect of reionization, we introduce a cosmic UV background (\citealp{Haardt2012}) at $z=7$ and linearly increase its strength until $z=6$, which is when cosmic reionization is believed to have completed (e.g., \citealp{Gunn1965, Fan2006}).

\subsection{Star formation}
As gas densities increase, stars are allowed to form as collisionless particles when the density surpasses a threshold of $n_{\rm H}=100$ $\rm cm^{-3}$, according to the Schmidt law (\citealp{Schmidt1959}). The star formation rate is governed by the equation $\dot{\rho}{\ast}=\rho_{\rm th}/\tau_{\ast}$, where $\tau_{\ast}=\tau_{\rm ff}/\epsilon_{\rm ff}$ is the star formation time scale. Here, $\tau_{\rm ff}=[3\pi/(32G\rho_{\rm th})]^{1/2}$ is the free-fall time at the threshold $\rho_{\rm th}$, and $\epsilon_{\rm ff}$ is the star formation efficiency per free-fall time. We set the star formation efficiency to $\epsilon_{\rm ff}\sim0.01$ for both Pop~III and Pop~II stars in this study, which is a typical value in the local Universe (e.g., \citealp{Leroy2008}).

The timescale for star formation is determined as follows:
\begin{equation}
\tau_{\ast}=\frac{\tau_{\rm ff} (n_{\rm H, th})}{\epsilon_{\rm ff}}\sim400 {\rm Myr} \left(\frac{n_{\rm H, th}}{100 \cmci}\right)^{-1/2}.
\end{equation}

 \subsubsection{Pop~III stars}
While the exact mass of metal-free stars in the early Universe remains uncertain, it is widely accepted that they were likely to be highly massive, as efficient cooling mechanisms were scarce at that time, with molecular hydrogen (H$_{\rm 2}$) being the primary coolant available (e.g., \citealp{Bromm2013, Whalen2013, Hirano2014, Stacy2016, Lazar2022}). Therefore, we adopt a top-heavy initial mass function (IMF) over a mass range of $[1-260]\msun$ for Pop~III stars to extract an individual star.
The functional form of the IMF is expressed as follows:
\begin{equation}
\phi = \frac{dN_{\rm PopIII}}{d \ln m} \propto m^{-1.3} \exp{\left[ -\left(\frac{m_{\rm char}}{m}\right)^{1.6}\right]}\mbox{\ ,}
\end{equation}
where $m_{\rm char}=60\msun$ is the characteristic mass.  

\subsubsection{Pop~II stars}
Pop~II stars form in gas clouds that have been contaminated by metals expelled by previous Pop~III SN. The threshold density for Pop~II star formation is the same as that for Pop~III stars, but there is an additional requirement that the gas metallicity should be higher than a critical metallicity value of $Z_{\rm crit}=10^{-5.5}\zsun$, which is motivated by dust-continuum cooling. Since the resolution of the gas mass is not high enough to directly convert a gas particle into an individual Pop~II star with a mass of around 1 $\msun$, the assumption is made that Pop~II stars form as a cluster with a total mass of 500 $\msun$, following the Salpeter initial mass function with a slope of $\alpha=1.35$ over the mass range of $[0.1-100]\msun$. Once a gas particle meets the two conditions of $n_{\rm th}$ and $Z_{\rm crit}$, it is replaced by a sink particle that accretes surrounding gas until it attains a mass of $M_{\rm Pop II}=500\msun$.

\subsection{Supernova feedback}
The fate of individual Pop~III stars depends on their initial masses, with stars in the $10-40\msun$ range undergoing core-collapse supernovae (CCSNe) with typical energies of $10^{51}$ erg and stars in the $140-260\msun$ range dying in pair-instability supernovae (PISNe) with energies of $10^{52}$ erg. We transfer the resulting SN energy as thermal energy to neighboring SPH particles, increasing their temperature and ejecting metals based on the properties of the SN progenitor. Pop~III metal yields provided by \citet{Heger2010} and \citet{Heger2002} are adopted for CCSNe/HNe, and PISNe, respectively, with seven metal species (C, N, O, Si, Mg, Ne, Fe) tracked for normal CCSNe and PISNe, and 16 metals tracked by adding nine species (Na, Al, Ca, Sc, Cr, Mn, Co, Ni, Zn) for HNe. 

The mass range at which Pop~III stars produce an HN is not yet fully understood (e.g., \citealp{Karlsson2013}). In this study, we aim to investigate the environmental conditions required for producing a star resembling AS0039 in a dwarf galaxy. For simplicity, whenever an HN event occurs, we assume a fixed metal yield from a $21\msun$ Pop~III progenitor (\citealp{Heger2010}), which is considered as a progenitor mass of AS0039. Varying metal yields based on progenitor mass would make it difficult to distinguish unique HN signatures. This, in turn, would hinder our ability to examine the conditions necessary for Pop~II stars to display HN signatures comparable to the observed AS0039. We note that each HN event releases $E_{\rm SN}=10^{52}$ erg of energy. For Pop~II stars, possible metal yields associated with various evolutionary phases, such as AGB, TypeII, and Type I~a SNe, are considered. \citet{Jeon2017} provides a detailed description of the models adopted for each event.

Metals expelled by SNe are spread among approximately 32 neighboring gas particles, $N_{\rm Ngb}\approx32$, located around the explosion site. This initial distribution results in the metallicity of $Z_{\rm i}$, which is determined by the metal mass divided by the number of neighboring gas particles and a spline kernel function $W(r)$, dependent on the distance from the metal ejection site. Afterward, the metals diffuse from the original gas particle to its surrounding particles through a diffusion process described by a diffusion equation.

\begin{equation}
\frac{dc}{dt} = \frac{1}{\rho} \nabla \cdot (D \nabla c)\mbox{\ ,}
\end{equation}, where $c$ is the concentration of a contaminant fluid per unit mass, and $D$ is the diffusion coefficient. To incorporate the diffusion process into SPH simulations, \citet{Greif2009} has discretized the equation for an particle $i$ as follows,
\begin{equation}
\frac{dc_i}{dt} = \Sigma_{j}K_{ij}(c_i-c_j)\mbox{\ ,}
\end{equation} where 
\begin{equation}
K_{ij} = \frac{m_j}{\rho_i \rho_j} \frac{4D_i D_j}{(D_i+D_j)} \frac{r_{ij} \cdot \nabla_i W_{ij}}{r^2_{ij}}.
\end{equation}
Here, the index $j$ indicates surrounding gas particles, $W_{ij}$ is the kernel, and $r_{ij}$ is the distance between particle $i$ and {j}.

\begin{table}
\centering
\begin{tabular}{c|c|c}
\hline
Name & ${Z_{\rm crit}}  [{Z_{\odot}}] $ & $f_{\rm HN}$ \cr
\hline
{\sc Z35-F50} & $10^{-3.5}$ & 50\% \cr
{\sc Z55-F50} & $10^{-5.5}$ & 50\% \cr
{\sc Z55-F100} & $10^{-5.5}$ & 100\% \cr
{\sc Z55-Early} & $10^{-5.5}$ & early 1 star \cr
{\sc Z55-Late} & $10^{-5.5}$ & late 1 star \cr
{\sc Z55-Mass} & $10^{-5.5}$ & mass range $25 - 40 \Msun$ \cr
\hline
\end{tabular}
\caption{Summary of the simulations. Column (1): Run name. Column (2): Critical metallicity for Pop~II star formation. Column (3): Fraction of Pop~III stars exploding as HNe.}
\label{tab:set}
\end{table}

\begin{figure}
  \centering
  \includegraphics[width=80mm]{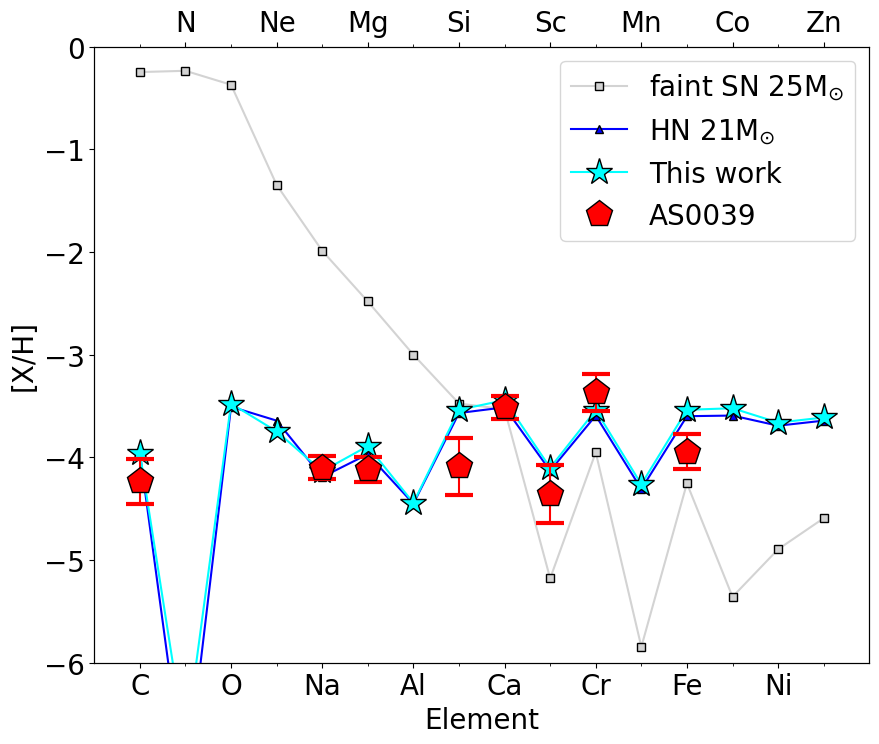}%
   \caption{The metal species of individual elements observed (represented by the red pentagon) are compared with the best-fit values provided by {\sc STARFIT} (shown as the blue line). This comparison suggests that the progenitor of the Pop~III star responsible for the observed metal abundances had a mass of $m_{\rm PopIII}\approx21\msun$, and it exploded as an HN explosion with an energy of $E_{\rm SN}=10^{\rm 52}$ erg. The abundance patterns produced by the HN explosion display unique characteristics that are different from those of a faint SN model, as depicted by the grey line. We also use {\sc STARFIT} to identify HN signatures in Pop~II stars generated in our simulated galaxies. An example of this is illustrated by the cyan line.
   }
   \label{fig:starfit}
\end{figure}

\begin{figure*}
      \centering
      \subfigure{\includegraphics[width=80mm]{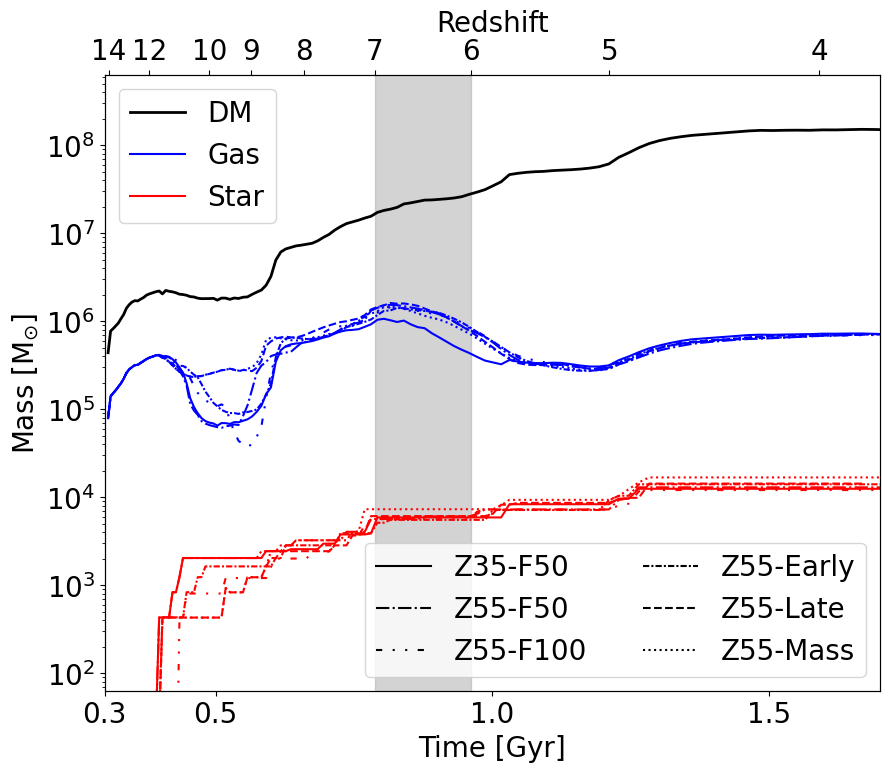}}
      \subfigure{\includegraphics[width=84mm]{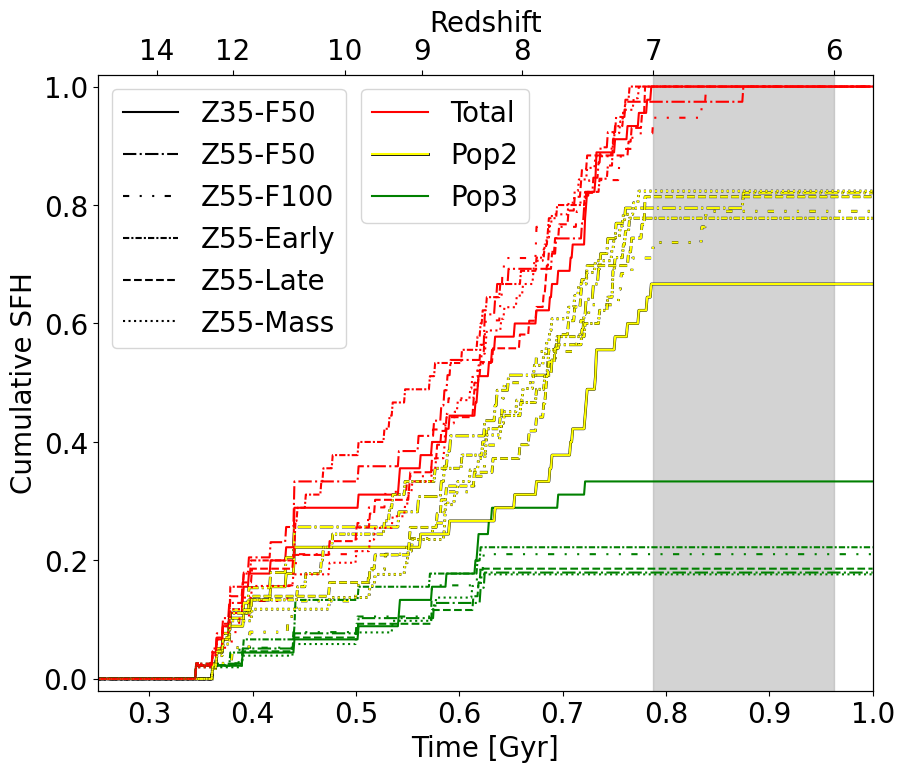}}
      \caption{The assembly histories of the simulated UFD analog are shown for six sets of simulations. The evolution of the masses for dark matter (black), gas (blue), and stars (red) is illustrated as a function of time in the left panel. The right panel displays the cumulative star formation histories for Pop~III (green) and Pop~II (yellow) stars, respectively. The formation of stars begins at approximately $z\approx12.9$ within minihaloes with masses of $M_{\rm vir}\approx10^{5-6}\msun$, which merged over time, eventually forming a galaxy with a mass of $M_{\rm vir}\approx3.8 \times 10^{8}\msun$ by $z=0$. We find that the star formation is halted by both cosmic reionization and SN feedback at $z=6.4$, resulting in a star formation period lasting approximately 550 Myr. It is crucial to emphasize that the emergence of the UFD analog at $z=0$ arises from the mergers of multiple progenitor haloes, indicating that the stars do not solely originate from the primary halo but also from other progenitor haloes. In the right panel, we show the history of stars originating from all progenitor haloes. In order to enhance comprehension regarding the distinction between stars formed in situ and those arising from external haloes, we explicitly differentiate these contributions in Figure \ref{fig:insitu_pop2}.}
      \label{fig:Mass_SFH}
  \end{figure*}

\subsection{Hypernovae}

\subsubsection{Sets of parameters}
To explore the circumstances that allow for the formation of a star similar to {\sc AS0039}, we have conducted simulations that altered two variables: the threshold metallicity required for the formation of Pop~II stars and the fraction of Pop~III stars that explode as HNe explosions. Table \ref{tab:set} presents information about the five different sets of simulations where we vary the two parameters mentioned previously.

\begin{itemize}
\item {\sc \bf Critical metallicity:}
The abundance of Pop~III stars and the occurrence of Pop~III HNe are determined by the critical metallicity. If the critical metallicity is low, mildly enriched gas can form Pop~II stars, whereas high critical metallicity could lead to the formation of Pop~III stars, increasing their abundance. There have been numerous studies aiming at narrowing down the value of critical metallicity (e.g., \citealp{Bromm2003, Santoro2006, Frebel2007, Dopcke2013, Chon2021, Sharda2022}). For instance, \citet{Bromm2003} suggests that fine-structure lines of carbon and oxygen dominate the transition from Pop~III to Pop~II, with a critical metallicity of $Z_{\rm crit} = 10^{-3.5}\zsun$. Alternatively, dust-induced fragmentation can also be a key driver for the formation of low-mass Pop~II stars, even at a low metallicity of $10^{-5.5}\zsun$ (e.g., \citealp{Tsuribe2006, Caffau2011, Schneider2012}). Since AS0039 has an extremely low stellar metallicity of $\rm [Fe/H]=-4.11$, we have chosen to use a critical metallicity of $Z_{\rm crit}= 10^{-5.5}$ $Z_{\odot}$ as a reference value, denoted as Z55 in the simulation names. We compare the results of those simulations with the run that uses a higher critical metallicity of $Z_{\rm crit}= 10^{-3.5}$ $Z_{\odot}$, as in the run named {\sc Z35-F50}.

\item{\sc \bf Frequency of hypernovae:}
The extent to which Pop~III stars undergo HN explosion at the end of their lives remains uncertain (e.g., \citealp{Umeda2005, Kobayashi2006, Nomoto2006, Yoon2012, Karlsson2013, Placco2015, Ishigaki2018}). To explore the full range of possibilities, the parameter $f_{\rm HN}$ is varied in this study. To establish the upper limit, it is assumed that all Pop~III stars experience HN explosion, designated as {\rm F100} in the simulation name. Conversely, the lower limit is set by allowing for a single HN explosion during the assembly history of the simulated UFD analog. Specifically, two cases are considered: one where the first Pop III star formed undergoes HN explosion ({\sc Z55-early}), and another where the single HN event happens randomly during the assembly process ({\sc Z55-Late}).

Based on the findings of \citet{Yoon2012}, which propose a possible mass range of $13\msun$ to $84\msun$ for HN events, about 50\% of Pop~III stars are expected to undergo HN events when using the IMF presented in equation (2). This fraction is also consistent with the study by \citet{Kobayashi2006}, which utilized simulations of galaxies similar to the Milky Way (MW) to explain observed chemical abundances, suggesting that to match the abundance of metal species such as $\rm [Zn/Fe]$, about half of the stars independent of mass and metallicity should explode as HNe, especially for low metallicity. In accordance with these findings, we adopt a value of $f_{\rm HN}$ of 50\%, labeled as F50 in the simulation names. We also take into account the work of \citet{Karlsson2013}, who propose that Pop~III stars in the mass range of $m_{\rm PopIII}=25-40\msun$ are capable of undergoing HN explosions, resulting in $f_{\rm HN}\approx11\%$. To incorporate their findings, we run a simulation called {\sc Z55-Mass}, in which HN explosions are allowed to occur among Pop~III stars in this mass range.

\subsubsection{Finding a hypernova signature}
We employ a tool called {\sc STARFIT} to extract the HN signatures from our simulations. This tool provides various metal yields that result from nucleosynthesis models (\citealp{Heger2002, Heger2010, Limongi2012, Just2015, Grimmett2018, Limongi2018}). We utilize these yields to determine several progenitor characteristics, such as mass, metallicity, and explosion energy. It is noteworthy that \citet{Skuladottir2021} also obtain the progenitor properties of AS0039 using {\sc STARFIT}, which are consistent with those of a primordial star having a mass of $m_{\rm PopIII}=21\msun$ and explosion energy of $E_{\rm SN}=10\times10^{51}$ erg, based on the metal abundances comparison. In Figure \ref{fig:starfit}, we present a comparison of individual metal abundances between AS0039 (depicted as a red pentagon) and the best-fit model generated from {\sc STARFIT} (represented as a blue line). We also show the abundances of Pop~II stars with HN signatures, obtained from our simulation, as a cyan line, which closely matches the result from {\sc STARFIT}. Additionally, we illustrate the metal species produced by a Pop~III star, which has an initial mass of $m_{\rm PopIII} = 25\msun$ and exploded as a faint supernova, as a gray line for comparison.

\end{itemize}

\section{Simulation results}
This section presents the findings of the simulated galaxy. Section 3.1 illustrates the process of mass assembly for the UFD analog. We then examine the number of progenitor minihaloes that could have Pop~II stars with Pop~III HN signatures and the fraction of such stars among all Pop~II stars in Section 3.2. Finally, we discuss the estimated chemical abundances of Pop~II stars from the simulations and compare them with observed values.

\subsection{Mass assembly}
Figure~\ref{fig:Mass_SFH} shows the mass assembly history of the simulated UFD galaxy analog (left panel) and the cumulative star formation history (right panel) by varying the fraction of Pop III stars that explode as HNe and the critical metallicity for Pop II star formation. In the left panel of Figure \ref{fig:Mass_SFH}, each color represents the dark matter (black), gas (blue), and stellar (red) masses. The gray-shaped region indicates the epoch of reionization, where the UV background is introduced and gradually increases in strength. Although the stochastic nature of star formation may result in slight differences in the gas and stellar mass during the assembly process, all sets of galaxies with the same initial conditions have grown into similar galaxies with a dark matter mass of $\rm M_{\rm DM}\approx 3.8 \times 10^8 \Msun$, a gas mass of $\rm M_{\rm gas}\approx 6.9 \times 10^5 \Msun$, and a stellar mass of $\rm M_{\ast}\approx 1.2 \times 10^4 \Msun$ at $z=0$, which are comparable to those of actual UFD galaxies. 

The simulated ratio of baryon mass to DM mass maintains a cosmological value of $f_{\rm b}=\Omega_{b}/\Omega_{m}\approx0.17$ up to a redshift of $z=11.5$, but it drops due to the feedback from Pop~III and Pop~II stars. Notably, as shown in the left panel of Figure~\ref{fig:Mass_SFH}, at around $\rm z \approx 10$, the baryon fractions in four runs ({\sc Z35-F50, Z55-F50, Z55-100}, and {\sc Z55-Early}) decrease to $f_{\rm b}=0.04$ due to the formation of Pop~II stars. These runs produce 2-5 times more stars than the other two runs ({\sc Z55-Late} and {\sc Z55-Mass}), leading to a reduction of the gas fraction. Furthermore, the emergence of the UV background causes a further decrease in the baryon fraction to below $f_{\rm b}\approx0.01$ at $z=6$, from $f_{\rm b}\approx0.08$ at $z=7$. Nonetheless, in all six sets of simulations, the target halo ultimately reaches a similar baryon fraction, even in the presence of different HNe frequencies. This outcome can be attributed to the relatively comparable total energy delivered by all SNe to the halo. Despite the significant impact of HNe, their events are surpassed in frequency by the higher occurrence of Pop~II SNe. The cumulative energy contribution from both SNe and HNe into the target halo, with respect to cosmic time, is illustrated in Figure~\ref{fig:SN_energy}.

\begin{figure*}
  \centering
  \includegraphics[width=180mm]{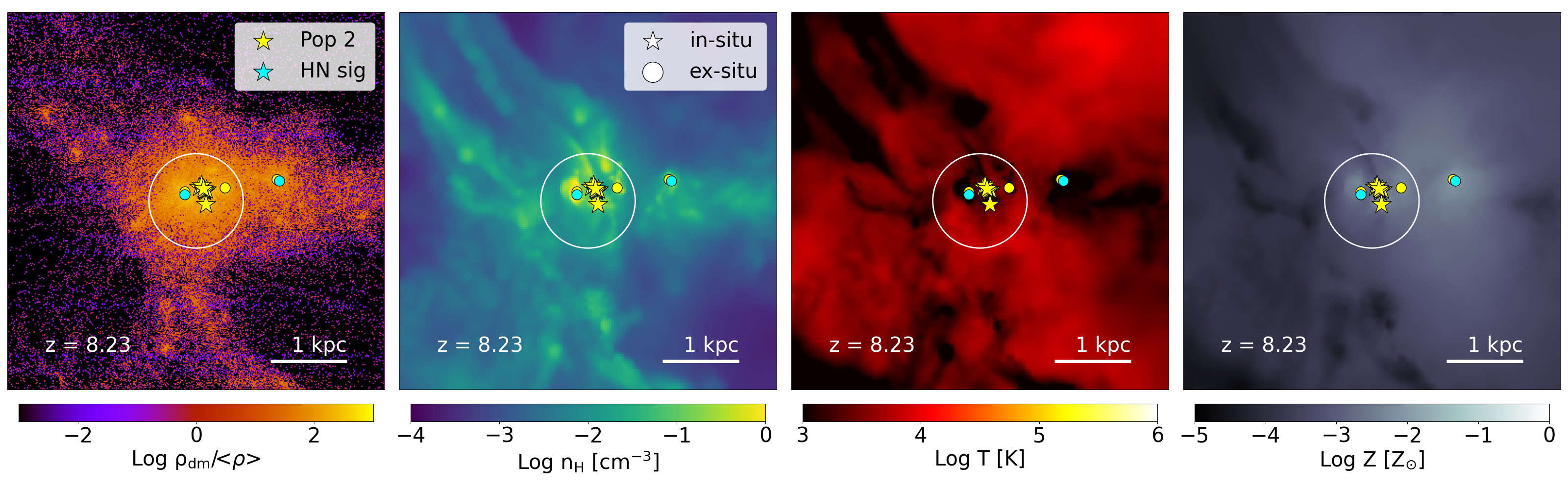}%
   \caption{The morphology of the simulated galaxy at $z\approx8.3$ is illustrated in four panels from left to right, displaying the dark matter overdensity, hydrogen number density, gas temperature, and gas metallicity projected along the line of sight within 1.75 kpc (4R$_{\rm vir}$) from the galaxy center. The virial radius of the halo is marked with solid white circles at the center of the panels. Pop~II stars with and without Pop~III HN signatures are represented by cyan and yellow colors, respectively. Moreover, in-situ stars and ex-situ stars are discerned through the use of star and circle symbols. These panels clearly demonstrate that stars are formed from multiple progenitor minihaloes and then merge into the primary halo at a later time.}
   \label{fig:morphology}
\end{figure*}

It is important to note that the formation of the UFD analog at $z=0$ is a result of small minihaloes merging together, indicating that it has multiple progenitor minihaloes. For reference, we offer a merger tree of the target halo in Appendix \ref{fig:mtree} for the {\sc Z55-F50} run. Our simulations reveal that in a small system such as UFD, the main progenitor does not dominate star formation at high redshifts, but rather, star formation is initiated in minihaloes of comparable mass, which merge later. It should be noted that we define a minihalo as a halo in which at least one Pop~III star has formed. The first star formation episode starts at around $z\approx12.9$ and ends at $z=6.5$, resulting in the star formation duration of approximately 550 Myrs. Therefore, the increase in stellar mass below $z=6$, depicted as the red lines in the left panel of Figure \ref{fig:Mass_SFH}, is due to the merging of progenitor minihaloes. It should be mentioned that we halt the simulations, except for {\sc Z55-F100}, when the redshift reached $z\approx3$ where the dark matter mass is $\rm M_{\rm DM}\approx 1.6 \times 10^8 \Msun$, the gas mass is $\rm M_{\rm gas}\approx 6.2 \times 10^5 \Msun$, and the stellar mass is $\rm M_{\ast}\approx 1.4 \times 10^4 \Msun$, respectively. At this point, star formation is completely quenched due to cosmic reionization, causing the maximum gas density inside the primary halo to drop to $n_{\rm H}\approx10^{-4}$ $\rm cm^{-3}$, and no further halo mergers occur. This was confirmed by running the {\sc Z55-F100} simulation down to $z=0$.

In order to explore the process of stellar assembly in more detail, we have included a plot of the cumulative star formation in the right panel of Figure \ref{fig:Mass_SFH}, which shows the star formation for Pop~III (green) and Pop~II (yellow) stars separately. Pop~III star formation takes place simultaneously in progenitor minihaloes of the simulated UFD analogs. For most simulations, Pop~III star formation ends around $z=8.5$. However, for the {\rm Z35-F50} simulation (solid line), which has a critical metallicity of $Z_{\rm crit}=10^{-3.5}~ Z_\odot$ compared to the other simulations with $Z_{\rm crit}=10^{-5.5}~Z_\odot$, Pop~III stars can be formed even in gas with higher metallicity. This causes the galaxy in the {\sc Z35-F50} simulation to continue forming Pop~III stars until around $z=7.5$, resulting in a higher stellar mass ratio occupied by Pop~III stars by about $10\%$ compared to the other simulations.

Our simulations show that the majority of minihaloes ($\sim83\%$) are likely to produce only a single Pop~III star before transitioning to Pop~II star formation. However, a small fraction of minihaloes is capable of producing multiple Pop~III stars, with 11\% having two and 3\% having three. It is worth noting that high-resolution simulations typically reveal the formation of Pop~IIIs in small groups (e.g., \citealp{Stacy2013, Stacy2016, Susa2019, Liu2021b, Jaura2022, Chiaki2022}). Therefore, given that we allow Pop~III stars to form as individual entities rather than as members of a stellar cluster composed of multiple stars, the feedback effect we present should be considered a lower limit. This is because the possibility of several sequential SN explosions might more effectively suppress subsequent star formation.

To investigate the role of merger events in the assembly of the simulated UFDs, we classify star formation for Pop~III and Pop II stars into in-situ and ex-situ star formation (see Table \ref{tab:insitu}). In-situ star formation refers to stars that are formed within the primary halo, which is defined as the halo contributing the most significant mass to the overall halo mass during the subsequent time step, especially when we trace back from $z=0$ to higher redshifts in search of the progenitor halo. Meanwhile, ex-situ star formation refers to stars that are formed in other progenitor haloes and then merge with the primary halo at a later time. Across all simulations, the {\sc Z35-F50} run has the highest total in-situ star formation for all stars, accounting for $\rm 26.7\%$, while the {\sc Z55-F100} run has the lowest value at $\rm 15.4\%$. The average in-situ star formation across all simulations is $\rm 19.5\%$, and approximately $\rm 80\%$ of the stars in the galaxies have grown through mergers. When comparing the in-situ ratio of Pop~III and Pop~II stars, the rate of Pop~II is higher by 8\%. This is because the duration of Pop~III star formation is shorter due to the transition from Pop III to Pop~II, and the halo mass becomes heavier when Pop~II star formation begins, leading to a higher in-situ rate compared to that of Pop~III stars.

The formation and merging of stars in multiple distinct minihaloes in the simulated galaxy at $z=8.3$ are depicted in Figure \ref{fig:morphology}. Each panel shows the projected dark matter overdensity, hydrogen number density, gas temperature, and gas metallicity within 1.75 kpc (4R$_{\rm vir}$) from the galaxy center, from left to right. Solid white circles at the center of each panel mark the virial radius of the halo. Different symbols represent different types of stars, with cyan stellar and yellow circle symbols representing Pop II stars with and without Pop III HN signatures, respectively. The figure illustrates that stars in the early Universe form in multiple progenitor minihaloes and then merge onto the primary halo at a later time. We find that the dispersal of gas observed in the gas density panel, particularly in relatively high-density clouds, is a consequence of an energetic HN explosion.

 \begin{table}
      \centering
      \begin{tabular}{c|c|c|c|c}
           \hline
           {Run} & {Total in-situ} & {Pop~III in-situ} & {Pop~II in-situ} \cr
           \hline
           {\sc Z35-F50} & 26.7\% & 20.0\% & 30.0\% \cr
           {\sc Z55-F50} & 23.1\% & 14.3\% & 25.0\% \cr
           {\sc Z55-F100} & 15.4\% & 11.1\% & 16.7\% \cr
           {\sc Z55-Early} & 15.6\% & 10.0\% & 17.1\% \cr
           {\sc Z55-Late} & 18.6\% & 12.5\% & 20.0\% \cr
           {\sc Z55-Mass} & 17.6\% & 11.1\% & 19.0\% \cr
           \hline 
      \end{tabular}
      \caption{The percentage of stars formed through in-situ star formation for each simulation set. Column (1) indicates the run name, column (2) represents the total in-situ fraction, column (3) shows the in-situ fraction for Pop~III stars, and column (4) indicates the in-situ fraction for Pop~II stars.}
      \label{tab:insitu}
  \end{table}

 \begin{table}
      \centering
      \begin{tabular}{c|c|c|c|c}
           \hline
            Run & $N_{\rm mini,Pop~III}$ & $N_{\rm mini,HN}$ & $N_{\rm mini,HN sig}$ \cr
           \hline
           {\sc Z35-F50} & 12 & 7 & 4  \cr
           {\sc Z55-F50} & 8 & 3 & 3  \cr
           {\sc Z55-F100} & 8 & 8 & 5  \cr
           {\sc Z55-Early} & 8 & 1 & 1  \cr
           {\sc Z55-Late} & 8 & 1 & 0  \cr
           {\sc Z55-Mass} & 8 & 1 & 1  \cr
           \hline
      \end{tabular}
      \caption{The number of progenitor minihaloes that merge to form a single UFD analog in each simulation set. Column (1) shows the simulation name, column (2) indicates the number of progenitor minihaloes that form at least one Pop~III star, column (3) shows the number of minihaloes with Pop~III stars that explode as an HN explosion, and column (4) indicates the number of minihaloes that contain Pop~II stars with inherited HN signatures.}
      \label{tab:num_halo}
  \end{table}

\begin{figure*} 
      \centering
      \includegraphics[width = 130mm]{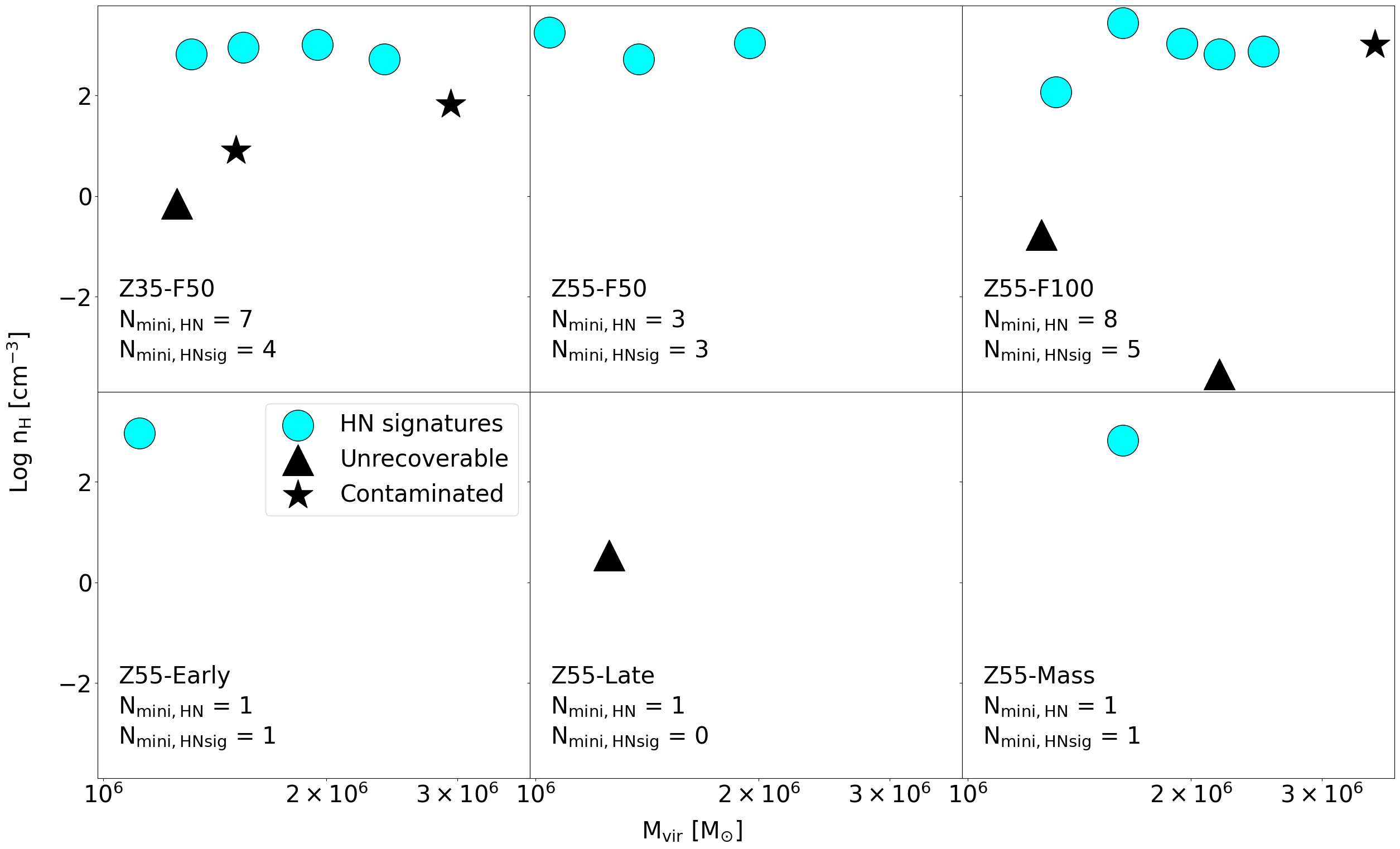}
      \caption{The virial mass of progenitor minihaloes and the number density of star-forming gas in those haloes for each simulation run. The minihaloes that contain Pop~II stars with the Pop~III HNe signatures are marked as filled cyan circles. Meanwhile, black symbols indicate minihaloes in which Pop~III HNe occurred, but no subsequent Pop~II star formation (indicated by triangles) or subsequent Pop~II stars are already contaminated by metals from other supernova events (represented by stellar symbols). The gas number density is determined immediately before the formation of Pop~II stars, but if the gas is unavailable due to significant evacuation, the gas density at about 70 Myr after the Pop III event is used. This time corresponds to the average time delay for the next star formation observed in the simulations.}
      \label{fig:n_H_minihalo}
\end{figure*}

\begin{figure} 
      \centering
      \includegraphics[width = 85mm]{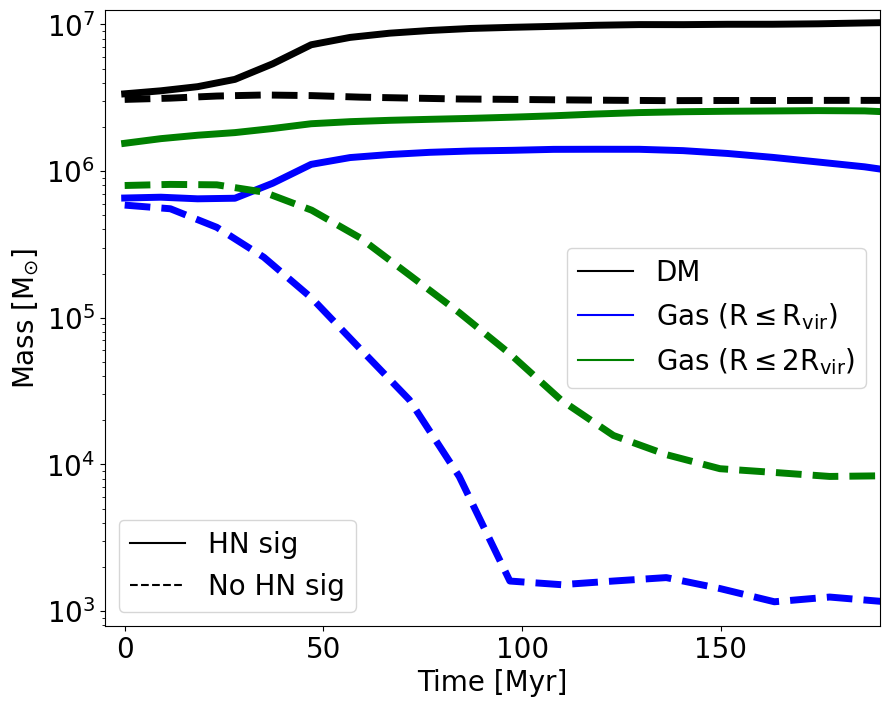}
      \caption{The evolution of two different haloes that experienced an HN explosion at similar virial masses. One of the minihaloes produces Pop~II stars with the Pop~III HN signature (solid line), while the other minihalo does not form any Pop~II stars (dashed line). The color of each line represents the DM mass (black), the gas mass within the viral radius, $R_{\rm vir}$ (blue), and the gas mass within $2 R_{\rm vir}$ (green). The density of star-forming gas in a rapidly growing dark matter halo (solid line) can be replenished by an influx of gas, while in a slowly growing dark matter halo (dashed line), it is difficult to recover gas that has evaporated into the IGM.}
      \label{fig:track_mass}
  \end{figure}
  
\subsection{Incidence of hypernova pattern}

We examine the percentage of progenitor minihaloes containing Pop~III stars that experience HN explosions and the number of minihaloes hosting the resulting Pop~II stars that inherit the Pop~III HN signatures within the same halo for each simulation. The corresponding numbers are summarized in Table \ref{tab:num_halo}. For example, in the {\sc Z35-F50} simulation, 12 different progenitor minihaloes have formed Pop~III stars, denoted as $N_{\rm mini,Pop~III}$=12. Of these, 7 minihaloes experience Pop~III HNe explosions, yielding $N_{\rm mini, HN}=7$. However, only 4 out of 7 progenitor minihaloes with HNe produce subsequent Pop~II stars with the Pop~III HN signature, noted as $N_{\rm mini, HNsig}=4$. Similarly, in {\sc Z55-F50}, Pop~III HNe events take place in 3 out of 8 progenitor minihaloes, and all of the subsequent Pop~II stars with the Pop~III HN signatures are found within these three minihaloes. On the other hand, if the frequency of Pop~III HNe is reduced, such as a single event, the number of minihaloes that could contain Pop~II stars with HN signatures also decreases.

Figure \ref{fig:n_H_minihalo} addresses the main question of the conditions that lead to the presence or absence of Pop~II stars with Pop~III HNe signatures. It shows the number density of star-forming gas within the minihaloes where an HN event is triggered as a function of their virial masses. Cyan circles on the figure correspond to minihaloes where Pop~II stars with Pop~III HNe signatures are present. On the other hand, the black symbols represent minihaloes where Pop~III HNe occurred, but either no subsequent Pop~II star formation (indicated by triangles) or the subsequent Pop~II stars are already polluted by metals from other supernova events (represented by stellar symbols). Regarding the value of number density, $n_{\rm H}$, it is chosen immediately before the formation of a Pop~II star. For other cases, the $\rm n_{H}$ value used is about $\sim$70 Myr after the Pop~III event, which corresponds to the average time delay for the next star formation observed in the simulations.

\begin{figure}
      \centering
      \includegraphics[width = 80mm]{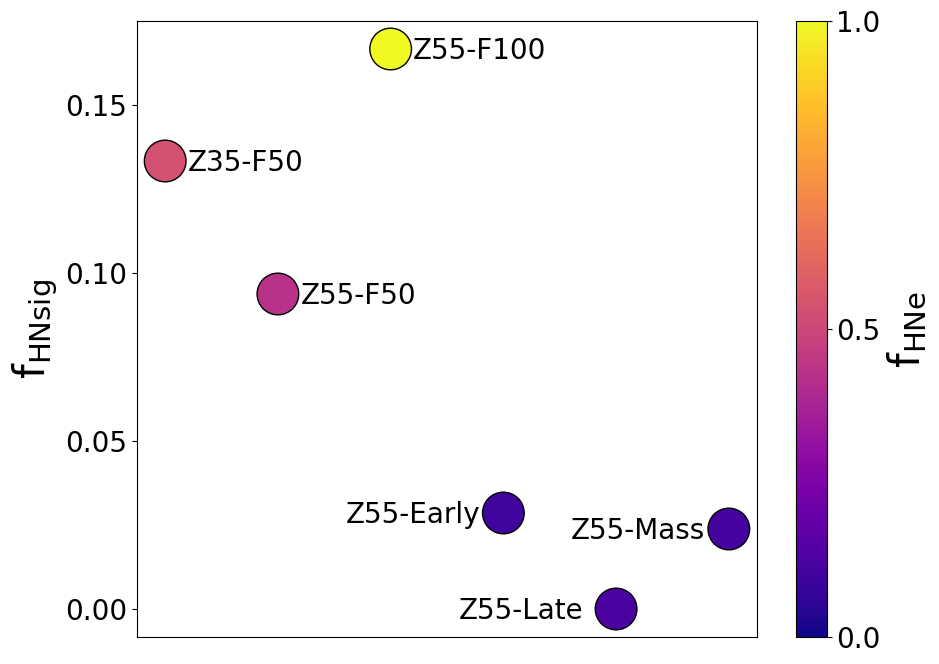}
      \caption{The fraction of Pop~II stars that exhibit HN signatures among all Pop~II stars in the simulated analog, denoted as $f_{\rm PopII, HN sig}$. As the fraction of Pop~III HNe decreases, there is generally a tendency for $f_{\rm PopII, HN sig}$ to show a declining pattern.}
      \label{fig:HNsig_fraction}
  \end{figure}

 \begin{figure*}
      \centering
      \includegraphics[width=130mm]{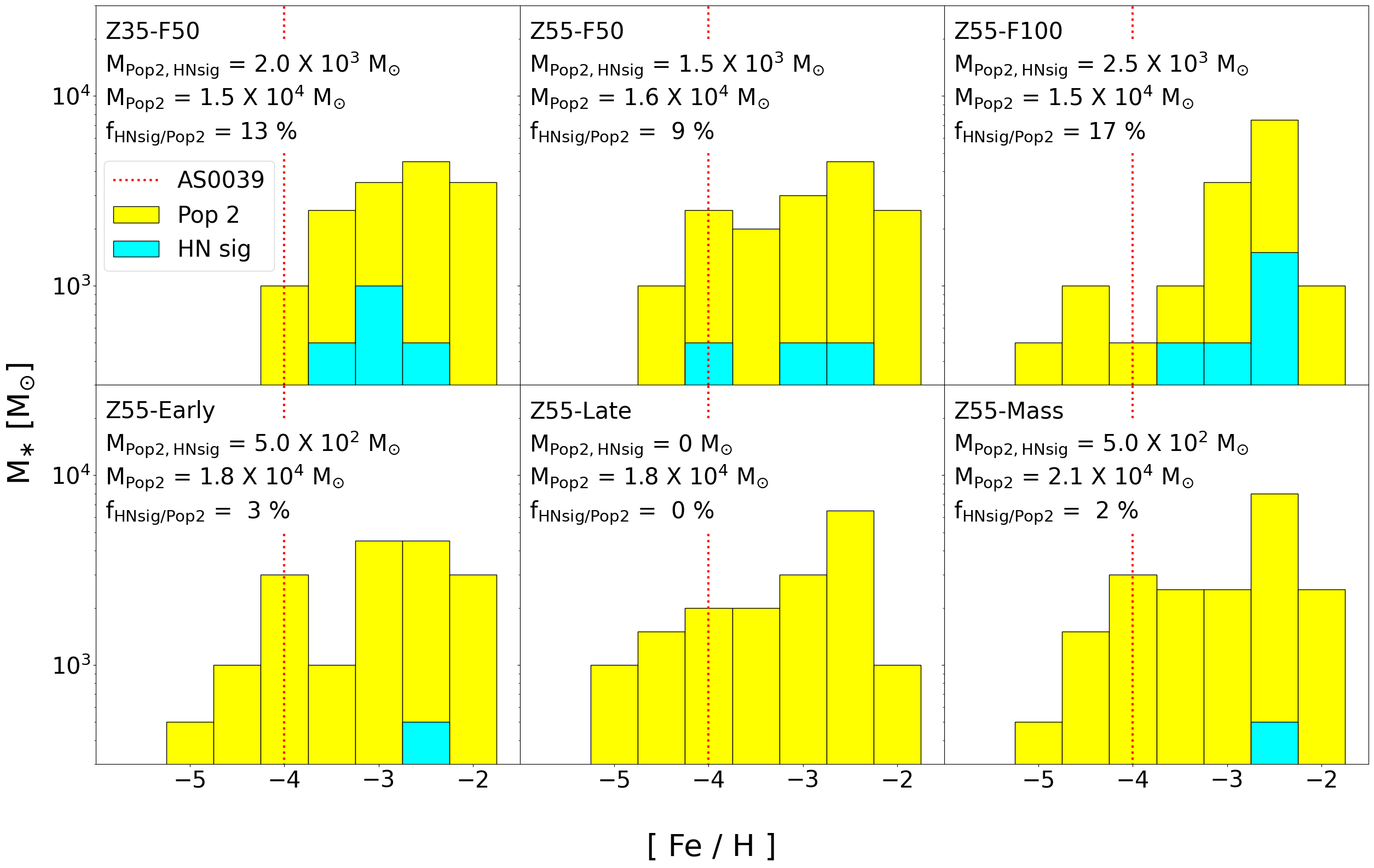}
      \caption{The metallicity distribution of Pop~II stars (yellow) found in the simulated UFD analog for each simulation run, comparing with Pop~II stars exhibiting the HN signatures (cyan). A vertical line is drawn to represent the metallicity of AS0039, which is $\rm [Fe/H]=-4.11$. With the exception of {\sc Z35-F55}, which is intended to prevent the formation of Pop~II stars with a metallicity lower than $10^{-3.5}\zsun$, most simulations result in approximately 18\% of UMP stars, defined as $\rm [Fe/H]<-4$.
      }
      \label{fig:pop2_hist}
  \end{figure*}

A total of 7 progenitor minihaloes, 3 from Z35-F50, 3 from Z55-F100, and 1 from {\sc Z55-Late} runs, are found to lack Pop~II stars with Pop~III HN signatures. These 7 cases can be divided into two categories. The first category, which we call the ``unrecoverable case," is characterized by a significant decrease in the number density of the star-forming region due to the strong feedback of HNe. As a result, the gas is unable to recover the threshold density of $n_{\rm H}=100$ $\rm cm^{-3}$, making further star formation impossible. We refer to the second category as the ``contaminated case". To form Pop~II stars with HN signatures, it is necessary that the gas is contaminated only by metals from HN events. However, in some scenarios, minihalo mergers occur before the formation of Pop~II stars, causing the gas to become polluted by metals from stars in the merged minihaloes. Consequently, Pop~II stars are formed out of gas with metals composed of those from Pop~III HNe and other SN events. The results of the study show that in the cyan-colored progenitor minihaloes, the gas is able to recover from Pop~III HN feedback and reach a gas density of $n_{\rm H} > 100$ $\rm cm^{-3}$. However, in the ``unrecoverable" cases, the gas density drops below $n_{\rm H}=100$ $\rm cm^{-3}$, with the Z55-F100 case being the most severely disrupted, where the density falls to $n_{\rm H}=0.001$ $\rm cm^{-3}$. In the ``contaminated" cases, the lack of Pop~II stars with HN signatures is caused not only by HN feedback but also by metal contamination resulting from mergers. Under these circumstances, densities of star-forming gas can be achieved up to $n_{\rm H}=10-1000$ $\rm cm^{-3}$.
  
Under what conditions can gas within a minihalo form Pop~II stars or become unrecoverable? It is likely that the surrounding environment plays a role in the observed difference. A comparison of two minihaloes with similar mass but different environments is presented in Figure \ref{fig:track_mass}. One minihalo produces Pop~II stars with the Pop~III HN signature (solid line), while the other minihalo does not form Pop~II stars (dashed line). The color of each line represents DM mass (black), gas mass within the viral radius (blue), and gas mass within two times the viral radius (green). When an HN explosion occurs with $E_{\rm SN}=10^{52}$ erg, the halo mass of both minihaloes is comparable, but the result differs due to the subsequent dark matter growth rate. In a fast-growing dark matter halo (solid line), the density of the star-forming gas is restored due to an influx of gas, whereas in a barely growing dark matter halo (dashed line), the gas that has evaporated into the IGM is difficult to recover. This is supported by the amount of gas within 2 $R_{\rm vir}$, which is about four times greater for the fast-growing minihalo than for the slow-growing halo around 100 Myr after the HN explosion event.

The question being asked is what fraction of Pop~II stars produced in the simulated UFD galaxy analog, which has a total stellar mass of approximately $10^4\msun$, contains Pop~III HN signatures. To address this, we identify the number of Pop~II stars with Pop~III HN signatures among all Pop~II stars within the viral radius of the simulated analog at $z=3$. Figure \ref{fig:HNsig_fraction} compares the fraction of Pop~II stars with HN signatures, represented as $f_{\rm PopII, HN sig}=N_{\rm PopII, HN sig}/N_{\rm PopII}$, in various runs. Except for {\sc Z35-F55} and {\sc Z55-LATE}, in general, $f_{\rm PopII, HN sig}$ tends to decrease as the fraction of Pop~III HNe decreases. Specifically, with the same critical metallicity of $Z_{\rm crit} = 10^{-5.5}$ $Z_{\rm \odot}$, the fraction is $f_{\rm PopII, HN, sig}=16\%$ for {\sc Z55-F100} and decreases to $f_{\rm PopII, HN, sig}=9.3\%$ for {\sc Z55-F50}, eventually dropping to $f_{\rm PopII, HN, sig}=2.3\%$ for {\sc Z55-Mass}.
  
\begin{figure}
      \centering
      \includegraphics[width = 85mm]{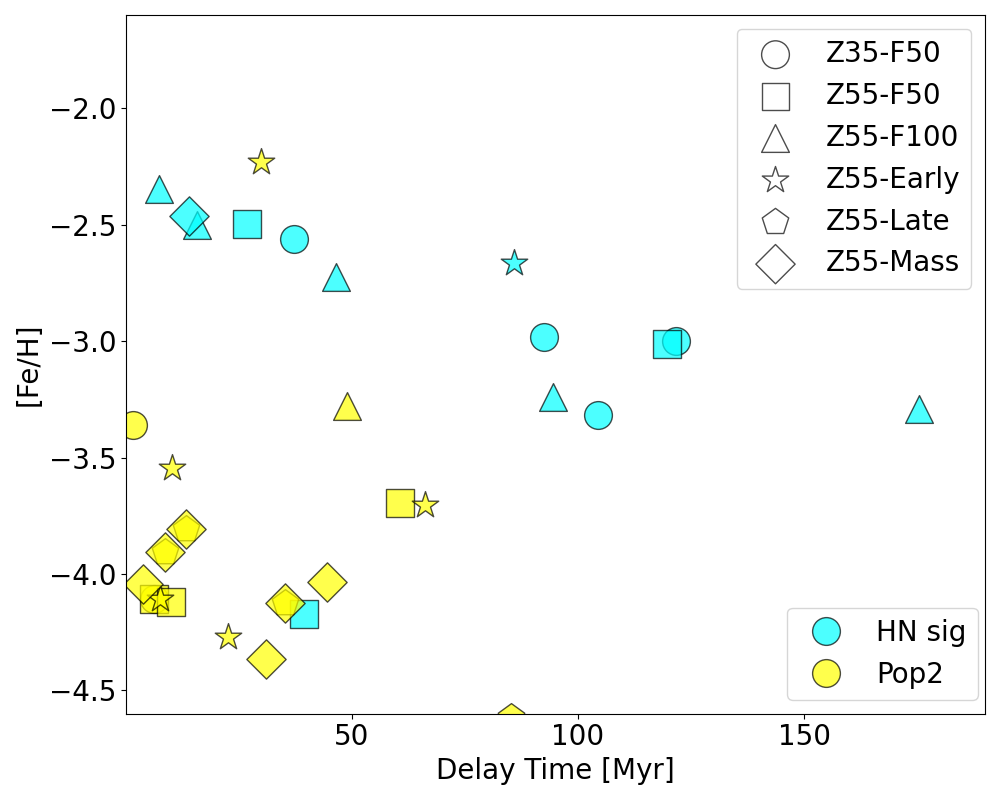}
      \caption{The metallicity of Pop~II stars, which formed right after the Pop~III star formation, is dependent on the delay time. The delay time is defined as the time gap between the Pop~III SN explosion and the formation of subsequent Pop~II stars. Note that the absolute amount of iron produced by a normal Pop~III SN is lower than that of an HN explosion from a Pop~III star with $m_{\rm PopIII}=21\msun$ (\citealp{Heger2010}), leading to lower overall [Fe/H] values. Pop~II stars that formed after a Pop~III HN explosion (cyan color) typically exhibit a longer delay time of up to about 180 Myr owing to the high-energy explosion, compared to those created after a normal SN (yellow). The reason for this is that metallicity tends to decrease with longer delay times, primarily due to increased diffusion unless there is external enrichment.}
      \label{fig:delay_time}
  \end{figure}

 \begin{figure}
      \centering
      \includegraphics[width = 85mm]{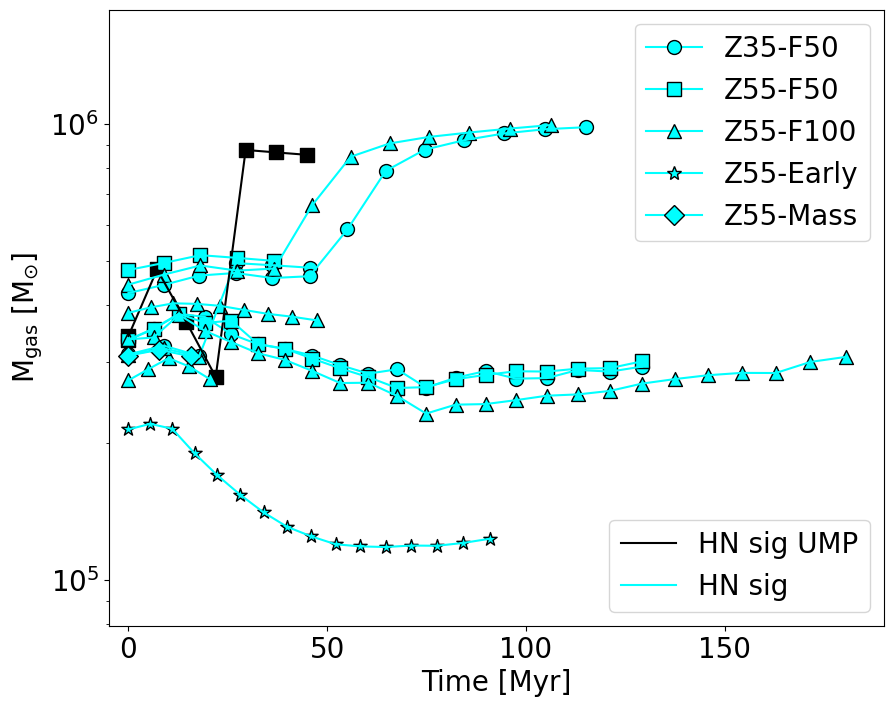}
      \caption{During the delay time period, the gas evolution within haloes plays a critical role in determining the metallicity of subsequently forming stars. In the case of the halo where UMP Pop~II stars are found (black line), there is a rapid gas inflow with a peak value of $\dot{M}_{\rm gas}\sim$ 0.08 $\msun \rm yr^{-1}$, which is significantly higher than the averaged gas inflow rate of $\sim$0.0025 $\msun \rm yr^{-1}$ for other haloes (cyan color). The significant increase in the infall rate observed in this case can be due to a merger event between the halo where the UMP star is formed and a main progenitor halo, taking place $\sim$25 Myr before the formation of the UMP star. This rapid influx of gas plays a crucial role in diluting the metals produced by Pop~III HN, thereby facilitating the formation of UMP stars.}
      \label{fig:gas_mass}
  \end{figure}
  
Assuming that half of the Pop~III stars explode as HNe, setting a higher critical metallicity of $Z_{\rm crit} = 10^{-3.5}$ $Z_{\rm \odot}$ in the {\sc Z35-F50} run results in a higher fraction of Pop~II stars with HN signatures, specifically with a value of $f_{\rm PopII, HN, sig}=13\%$. This is higher than the $f_{\rm PopII, HN, sig}=9.3\%$ observed in the {\sc Z55-F50} run. The reason for this is that in the {\sc Z35-F50} run, there are more Pop~III stars, leading to a higher frequency of HN events. In fact, HN events occur 2.6 times more frequently in the {\sc Z35-F50} run than in the {\sc Z55-F50} run. Out of the 6 sets considered, the runs that have only one HN event ({\sc Z55-Early}, {\sc Z55-Late}, {\sc Z55-Mass}) all exhibit a $f_{\rm PopII, HN sig}$ that is lower than $3\%$. It should be noted that the $f_{\rm PopII, HN sig}$ predicted above are likely to be lower in more massive satellite galaxies than in UFDs. This is because larger galaxies are expected to have a higher number of Pop~II stars. Although the number of Pop~III stars exploding as HNe may also increase with halo mass, the duration of Pop~III star formation in the early universe is very short $z>7.5$, making Pop~II stars the dominant population in more massive galaxies. As a result, the fraction of Pop~II stars with HN signatures compared to the total Pop~II stars is likely to decrease as the halo mass increases.

\begin{figure*}
      \centering
      \includegraphics[width = 150mm]{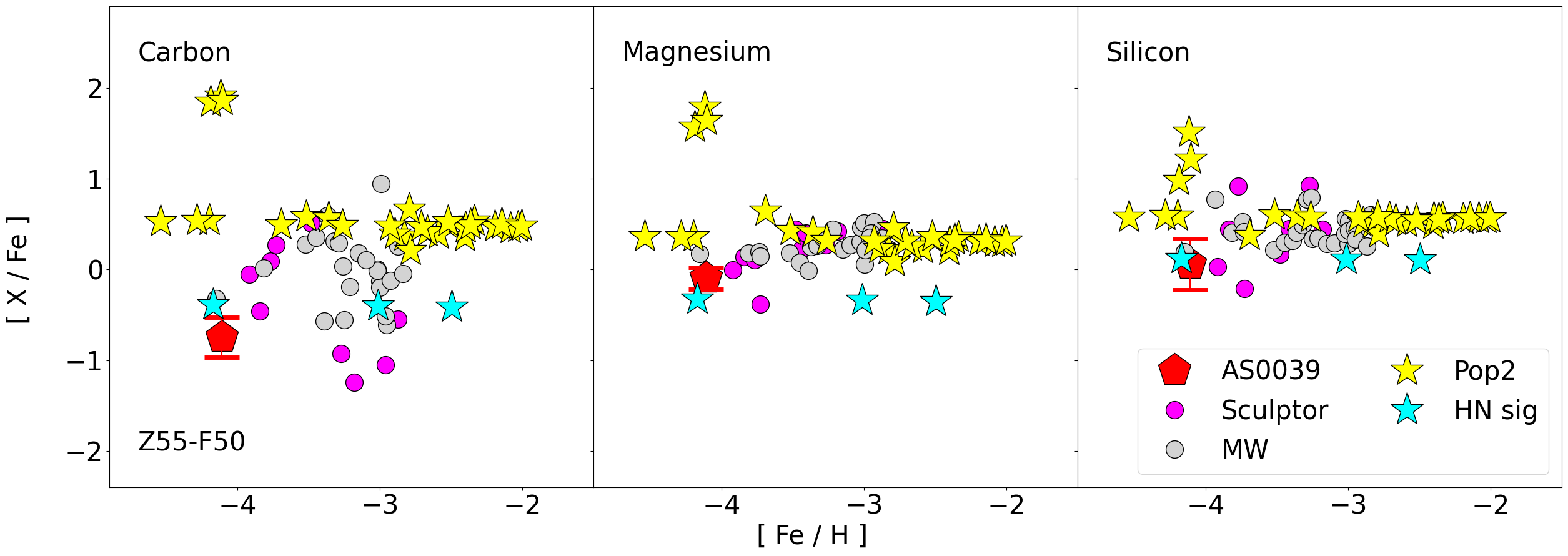}
      \caption{Comparison of individual metal abundances between our simulation of {\sc Z55-F50} (stellar symbols) and observations. The observations include AS0039 (red pentagon, \citealp{Skuladottir2021}), stars in the Sculptor galaxy (magenta circle, \citealp{Frebel2010, Simon2010, Tafelmeyer2010, Starkenburg2013, Jablonka2015}), and stars in the MW (grey circle, \citealp{Cayrel2004}). Pop~II stars with HN signatures (cyan) show [X/Fe] values that are lower than those of normal Pop~II stars (yellow) by $\sim$1 dex for carbon and by 0.5-0.7 dex for silicon and magnesium. The abundances of Pop~II with HN signatures are in good agreement with those of {\sc AS0039}. Our simulations demonstrate that the difference in carbon abundance is due to the type of SN explosion. Specifically, Pop~II stars that form out of gas contaminated by a normal Pop~III SN with low iron abundance tend to give rise to CEMP stars with $\rm [C/Fe]\approx2$. In contrast, the low carbon value of $\rm [C/Fe]=-0.5$, which matches the value observed in AS0039, is attributed to a Pop III HN explosion. Moreover, we find that Pop~II stars with $\rm [C/Fe]\approx0.6$ are more affected by Pop~II SNe than Pop~III SNe.}
      \label{fig:C_Mg_Si}
  \end{figure*}
  
\subsection{Chemical abundances}
Figure \ref{fig:pop2_hist} shows a comparison between the metallicity distribution of all Pop~II stars (yellow) and Pop~II stars with HN signatures (cyan) found in the viral radius of the UFD analog at $z=3$ for each run. In all runs except {\sc Z35-F50}, where the formation of Pop~II stars with metallicity below $Z_{\rm crit}=10^{-3.5}$ $Z_{\odot}$ is not possible, the metallicity of the Pop~II stars ranges from [Fe/H] $\approx-5.5$ to [Fe/H] $\approx-1.5$ (yellow). The vertical line in Figure \ref{fig:pop2_hist} represents the stellar metallicity of {\sc AS0039}, which is measured to be $\rm [Fe/H]=-4.11$, corresponding to the ultra metal-poor (UMP) stars ($\rm [Fe/H]<-4$) (e.g., \citealp{Beers2005}). When all runs are combined, the fraction of UMP stars is around 18\% out of a total of 204 Pop~II stars, with 37 of these stars being UMP stars. As such, the fraction of UMP stars among only the Pop~II stars with HN signatures (cyan) is $\sim$7\%, which is based on a total of 14 Pop~II stars with HN signatures, among which only 1 star is a UMP star. The Pop~II star with the HN signature that has the most similar metallicity to AS0039 is the one with [Fe/H]=-4.17 in the Z55-F50 run. Most of the other Pop~II stars with HN signatures have metallicities in the range of -3.4 to -2.4 for their [Fe/H] values.

The run {\sc Z55-early} is the simplest and most suitable run to investigate the conditions for having such a low metallicity like the observed AS0039 in that only the first Pop~III star in the first minihalo explodes as an HN. It still, however, leads to the formation of a Pop~II star with an HN signature, albeit with a metallicity, $\rm [Fe/H]\approx-2.6$, more than one order of magnitude higher than that of AS0039. Comparably, when a Pop~III HN explodes in a minihalo that collapses relatively late, like in the runs {\sc Z55-late} or {\sc Z55-Mass}, even if it is a single event, the HN unique signatures can be easily diluted by metals from other SN events (e.g. \citealp{Ji2015}). As a result, none of the Pop~II stars with HN signature can be formed in the {\sc Z55-late} run owing to external pollution. However, if subsequent star formation occurs before the gas is contaminated, the unique signature remains, forming Pop~II stars with it, as in the {\sc Z55-Mass} run.

According to our findings, the metallicity of Pop~II stars is related to the delay time between the SN explosion and subsequent Pop~II star formation. Figure \ref{fig:delay_time} illustrates this correlation by plotting stellar metallicity against the delay time, which is the interval between the occurrence of a Pop~III SN explosion and the formation of Pop~II stars from the gas contaminated by the SN event. In general, Pop~II stars with HN signatures (cyan color) tend to have higher [Fe/H] values than those formed after a normal Pop~III SN (yellow). This is due to the fact that the absolute yield of iron from a normal Pop~III SN with an energy of $\rm 10^{51}$ erg is lower than that of an HN explosion from a progenitor mass of $m_{\rm PopIII}=21\msun$ (\citealp{Heger2010}). Additionally, the high energy associated with HN explosions leads to longer delay times, up to $\sim$180 Myr, for Pop~II stars with HN signatures. Our simulation results clearly show that the metallicity of Pop~II stars with HN signatures decreases with longer delay times. For instance, Pop~II stars with a metallicity of $\rm [Fe/H]\approx-2.5$ require a delay time of about $\sim$40 Myr, while the formation of EMP stars needs a delay time of about $\sim$100 Myr. This is because the metals produced by the HN event would have been diluted during the long delay time period, leading to a decrease in the gas metallicity where the Pop~II stars are formed. This effect becomes more pronounced as the delay time increases, and without any external metal pollution during this period, the subsequent Pop~II stars will form with low metallicities.

\par 
Although Pop~II stars with HN signatures exhibit [Fe/H] values above $\rm [Fe/H]\approx-3.3$, we have only generated one UMP star with $\rm [Fe/H]\approx-4.17$, similar to that of AS0039. To investigate the conditions that lead to the formation of such UMP stars, we analyze the gas evolution within the halo in which Pop~II stars with HN signatures reside during the delay time period, as shown in Figure \ref{fig:gas_mass}. In contrast to other haloes (cyan color), the halo where the UMP star, presented as the black line, is formed experiences a rapid gas inflow of about $M_{\rm gas}\approx5\times10^5\msun$ within $\sim$7 Myr, about 22 Myr after the HN explosion, increasing the gas mass from $M_{\rm gas}\approx2.8\times10^5\msun$ to $M_{\rm gas}\approx8.8\times10^5\msun$. The resulting gas accretion rate is $\dot{M}_{\rm gas}\sim$ 0.08 $\msun \rm yr^{-1}$, which is 4 times higher than the average gas accretion rate onto the halo. This peak value is higher by a factor of 30 compared to the average gas inflow rate of $\dot{M}_{\rm gas}\sim0.0025\msun \rm yr^{-1}$ for other haloes. The observed high infall rate, in this case, can be attributed to a merger event occurring $\sim$25 Myr prior to the formation of the UMP star. This merger involves the combination of a halo hosting a UMP star similar to AS0039 with a main progenitor halo, resulting in the influx of $M_{\rm gas}\sim5\times10^5\msun$ of gas. The rapid gas influx into the halo where the UMP star forms plays a crucial role in diluting the metals produced by the Pop~III HN, ultimately allowing the formation of UMP stars. Consequently, our findings suggest that low-metallicity stars can form via two possible mechanisms: a prolonged delay time or a substantial inflow of gas onto a halo, which enables the diffusion of metals.

\begin{table*}
\centering
\begin{tabular}{ c | c| c| c}
\hline
Parameter & Description & Best Fit  & reference  \\
\hline
$v_{\rm SV}$ & Baryonic streaming velocity & $0.8 \sigma_{\rm SV}$ & (1)   \\
\hline
$M_{\rm min}$  & Minimum mass of Pop~III stars &  $5\msun$  & (1)  \\
\hline
$M_{\rm max}$  & Maximum mass of Pop~III stars &  $210\msun$  & (1)  \\
\hline
${\rm \eta_{III}}$  & Star formation efficiency for Pop~III stars &  0.38   & (1) \\
\hline
${\rm \eta_{II}}$  & Star formation efficiency for Pop~II stars &  2   & (2) \\
\hline
$\alpha_{\rm out}$  & Slope of outflow efficiency &  0.72   & (2) \\
\hline
$M_{\rm out,norm}$  & Normalization mass of outflow efficiency &  $10^{10.5}\msun$   & (2) \\
\hline
\end{tabular}
\caption{Best-fit parameters for {\sc A-SLOTH} we adopt for this work. Column 1): Parameters. Column 2): A short description of the parameters.
Column 3): Values. Column 4): References for the adopted values are as follows: (1) \citet{Hartwig2022}, and (2) \citet{Chen2022}.}
\label{tab:asloth_param}
\end{table*}

It should be mentioned that \citet{Jeon2014} suggested a longer recovery timescale of up to $\sim$300 Myr when using the energy of $E_{\rm SN}=10^{52}$ erg for a Pop III SN. Such difference in timescale compared to our work can be attributed to two factors. Firstly, \citet{Jeon2014} examined halo masses of $M_{\rm vir}\approx5\times10^5\msun$, which are $2-10$ times smaller than the masses used in our study to estimate the recovery time. Secondly, we do not account for the pre-explosion photoionization by the Pop~III progenitors, which may have caused gas densities to remain high prior to the SN explosion, resulting in a shorter recovery timescale in our simulations. Even when considering the photoionization effect, however, recovery time can vary considerably depending on the halo mass, as illustrated by \citet{Jeon2014}. For example, the recovery time for a $40\msun$ Pop~III progenitor star that explodes with $E_{\rm SN}=10^{51}$ erg was $\sim$92 Myr for a halo mass of $M_{\rm vir}=5\times10^5\msun$, but dramatically dropped to $\sim$6 Myr for a halo mass of $M_{\rm vir}=9\times10^5\msun$. This is due to the H~II region becoming more compact, which weakens the pre-heating effect and allows gas to rapidly recollapse in relatively massive haloes.

In Figure \ref{fig:C_Mg_Si}, we compare the individual abundances of Pop~II stars with HN signatures (cyan star) and normal Pop~II stars (yellow star) from our simulations to observations of stars, such as AS0039 (red pentagon, \citealp{Skuladottir2021}), stars in the Sculptor galaxy (magenta circle, \citealp{Frebel2010, Simon2010, Tafelmeyer2010, Starkenburg2013, Jablonka2015}), and stars in the MW (grey circle, \citealp{Cayrel2004}). The focus of the comparison is on the abundances of Carbon, Magnesium, and Silicon, which are displayed in three panels in Figure \ref{fig:C_Mg_Si}. Our study reveals that the individual abundances of Pop~II stars with HN signatures, depicted as a cyan star, are lower than those of normal Pop~II stars (yellow star), ranging from -1 dex for Carbon to -0.5 dex for Magnesium and Silicon. The estimated abundances of Pop~II stars with HN signatures are consistent with the best-fit values of AS0039, given the uncertainties in the observations. Note that the abundances of AS0039 are measured values, and there could be differences between the observed and the nucleosynthesis results, even with the best-fit model.

It is worth noting that AS0039 has a peculiar feature in its carbon abundance; unlike CEMP stars commonly found in dwarf galaxies, which are also considered signatures of Pop~III stars, AS0039 is a carbon-poor star with $\rm [C/Fe]_{LTE} = -0.75 \pm 0.22$. Our simulations clearly exhibit that this difference in carbon abundance is attributed to the type of Pop III SN. Specifically, Pop II stars with an exceptionally high [C/Fe] ratio of $\sim$2 tend to form when Pop II stars arise immediately after a normal SN explosion with low iron abundance, leading to the formation of CEMP stars. In contrast, the low value of $\rm [C/Fe]\approx-0.46$, which matches the observed value of AS0039, is the result of the HN explosion (cyan). Furthermore, Pop~II stars with $\rm [C/Fe]\approx0.6$, which form in gas clouds that are more influenced by Pop~II than Pop~III stars, do not exhibit CEMP features. However, these stars still have distinct features that differ from those with HN signatures, as shown in AS0039.

We observe the narrow scatter of stellar abundances in normal Pop~II stars showing similar $\rm [X/Fe]$ values compared to those of stars in Sculptor and the MW. This similarity is likely the result of the short star formation duration of $\sim$0.5 Gyr in the simulated UFD analog, while Sculptor has had a star formation duration of $\rm \Delta$t $\approx$ 1-3 Gyr, as reported in previous studies (\citealp{Kirby2011, Weisz2014, Bettinelli2019, delosReyes2022}), and the MW is undergoing ongoing star formation (\citealp{Elia2022}).


\section{Observability}
So far, we have discussed the simulation results for a single dwarf galaxy with a halo mass of $M_{\rm vir}\approx10^8\msun$ at $z=0$, by varying the fraction of Pop~III stars that undergo HN explosions. However, the important question is to determine the likelihood of observing a Pop~III HN signature in a randomly observed dwarf satellite galaxy within the MW's volume and which galaxy masses are most likely to exhibit such signatures. To address these questions, we employ a semi-analytic approach to estimate the number of 
progenitor minihaloes that are expected to contain Pop~III HN remnants within the MW's volume.

\subsection{{\sc A-SLOTH}: semi-analytic model for galaxy formation}
We utilize a semi-analytic galaxy formation code called {\sc A-SLOTH} (Ancient Stars and Local Observables by Tracing haloes, \citealp{Hartwig2022, Magg2022}) to track the remnants of early generations of stars down to $z=0$ (e.g., \citealp{Magg2018, Hartwig2019, Chen2022}). This approach involves using the Extended Press-Schechter (EPS) formalism (e.g., \citealp{Press1974, Lacey1993}) to generate halo merger trees for an MW-like galaxy or obtaining them from N-body simulations. Particularly, {\sc A-SLOTH} adopts the halo merger tree extracted from the {\sc Caterpillar} project (\citealp{Griffen2016}), which provides 30 sets of merger trees for MW-sized galaxies using dark matter-only simulations. By traversing through these merger trees, {\sc A-SLOTH} is able to ascertain the baryonic components of each halo in the tree based on the implemented physics within the code, starting from the first star formation in minihaloes.

One of the key benefits of using a semi-analytic approach is its computational efficiency, which allows for rapid exploration of the optimal parameters that describe baryonic physics. Further information on the specific baryonic physics incorporated in {\sc A-SLOTH} can be found in \citet{Hartwig2022}. Compared to other semi-analytic models, {\sc A-SLOTH} stands out by including the relevant physics of early star formation, allowing for the formation of individual Pop~III and massive Pop~II stars, and taking into account the impact of their mechanical and chemical feedback.

The models are calibrated based on six distinct observables to determine the best-fit values of the free parameters that govern baryonic physics. These include the optical depth to Thomson scattering, the stellar mass of the MW, the cosmic star formation rates at high redshift, the distribution of stellar masses among the MW's satellite galaxies, the fraction of extremely metal-poor stars (EMP) (i.e., [Fe/H]$<$-3) in the halo, and the ratios of UMP (i.e., [Fe/H]$<$-4) to EMP stars. For this study, we adopt not only the proposed values of the free parameters by \citet{Hartwig2022} but also the values by \citet{Chen2022}, where they acquire the best-fit values by focusing on the relationship between the stellar mass and halo mass and adjusting the Pop~II star formation models to match observed values. Table \ref{tab:asloth_param} summarizes the parameters for {\sc A-SLOTH} that we use in this work.

\begin{figure}
  \centering
  \includegraphics[width=90mm]{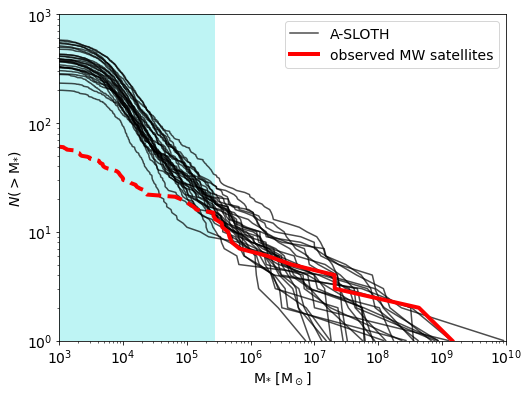}%
   \caption{The cumulative number of satellite galaxies obtained from 30 sets of merger trees for the MW-like galaxies using {\sc A-SLOTH} (black lines), with a red line indicating observational data for comparison (\citealp{McConnachie2012, Munoz2018}). It is worth noting that the blue-shaded region refers to galaxies with stellar masses below  $2.9\times10^5 M_{\odot}$, which are incomplete in observations. Although {\sc A-SLOTH} is calibrated based on various physical quantities, such as the mass of the MW galaxy at $z=0$ and global star formation histories, it is likely to predict a higher number of low-mass galaxies ($M_{\ast} \lesssim 2\times10^4 M_{\odot}$) by a factor of 2-10 compared to what is observed.}
   \label{fig:Cum_num_sat}
\end{figure}

In Figure \ref{fig:Cum_num_sat}, we show the cumulative number of satellite galaxies resulting from 30 sets of merger trees of the MW-like galaxy using {\sc A-SLOTH}, indicated by the black lines, compared to the observational data represented by the red line (\citealp{McConnachie2012, Munoz2018}). It is important to note that the observations of galaxies with stellar masses below $M_{\ast} \approx2 \times10^5\msun$, which are represented by the blue-shaded region, are still incomplete (e.g., \citealp{Carlsten2021}). The mean stellar mass of the MW-like galaxy, calculated by averaging over 30 sets of {\sc A-SLOTH} simulations, is approximately $M_{\ast}\approx5.3\times10^{10}\msun$, which is consistent with the observed stellar mass of the MW galaxy of $4.86-6\times10^{10}\msun$ (e.g., \citealp{McMillan2017}). The {\sc A-SLOTH} simulation, however, is likely to predict a large amount of low-mass satellite galaxies ($M_{\ast}\lesssim 2\times 10^4\msun$) that exceed what is actually observed by a factor of 2-10. This discrepancy, commonly known as the ``missing satellite problem", (e.g., \citealp{Moore1999, Klypin1999, Bullock2017}) may indicate that there are numerous low-mass galaxies that remain undiscovered due to observational limitations or that the stellar feedback implemented in the simulation for small galaxies is not strong enough to suppress star formation. The following sections discuss the likelihood of detecting Pop~III remnants with HN signatures in two cases. One case will account for the difference between observations and {\sc A-SLOTH} simulations, while the other will not consider this discrepancy.

\subsection{The effect of Pop~III hypernovae}
We have employed {\sc A-SLOTH} to examine the prevalence of Pop~III HN explosion signatures in the MW dwarf satellites. We have experimented with the fraction of Pop III stars that die as HN explosions, defined as $f_{\rm HN}$, in five different scenarios: 1) all Pop~III stars end up as HNe, $f_{\rm HN}$=100\%, as in our {\sc Z55-F100}; 2) 50\% of Pop III stars die as HNe, comparable to our {\sc Z35-F50} and {\sc Z55-F50}; 3) a small proportion of Pop~III stars (either 5\% or 1\%) expire as HNe, which we compare to {\sc Z55-early} and {\sc Z55-late}; 4) Pop III stars having stellar mass from $25\msun$ to $40\msun$ end as HNe, in line with {\sc Z55-Mass}. Note that a Pop~III HN event is triggered with an explosion energy of $10^{52}$ ergs. During each simulation, we identify progenitor minihaloes that are considered to preserve Pop~III HNe signatures within the satellite galaxies of the MW-like galaxy.

minihaloes that exhibit HN signatures are defined as those containing Pop~II stars, which are formed from gas contaminated by Pop~III HNe via internal or external enrichment. If a Pop~III star undergoes an HN explosion within a minihalo, it is assumed that subsequent Pop~II stars born immediately after the explosion may display an HN signature on their surface. In the case of external enrichment, Pop~II stars are formed without prior Pop~III star formation but are contaminated with HN metals originating from an HN event in an adjacent minihalo. It is crucial to note that we only consider minihaloes with Pop~II stars formed immediately after contamination with a Pop~III HN explosion, as our simulations have shown that only these stars can maintain HN signatures. Otherwise, the unique signatures are likely to be mixed and washed out by metals from other SN events.

\begin{figure}
  \centering
  \includegraphics[width=90mm]{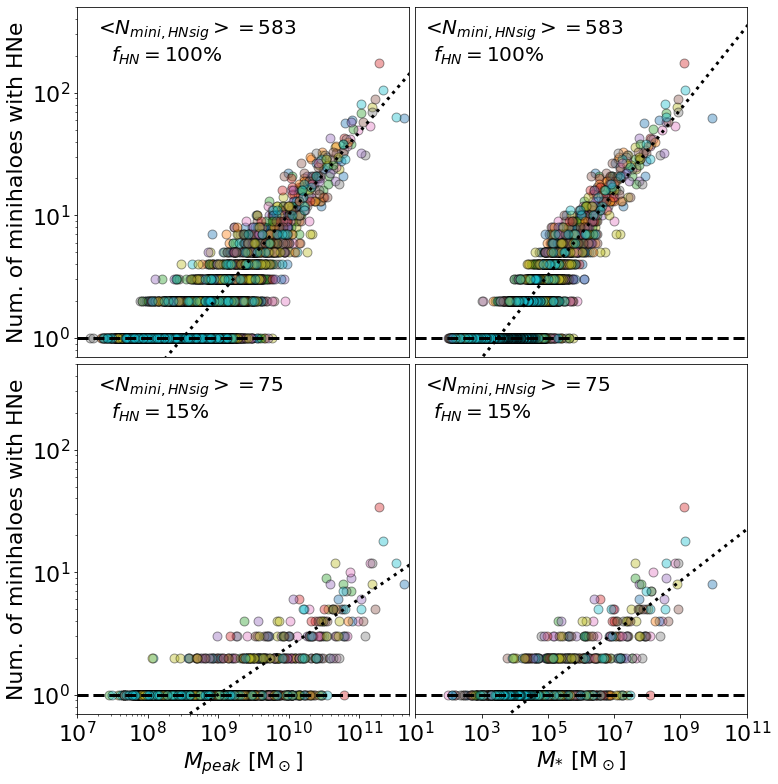}%
   \caption{The number of progenitor minihaloes that have undergone at least one Pop III HNe explosion in an MW-like galaxy, with respect to halo peak mass (left panels) and stellar mass (right panels). It is evident that more massive haloes tend to harbor a greater number of minihaloes undergoing Pop~III HNe events, as clearly depicted by the linear fit represented as a dotted line. Additionally, we mark the halo that has undergone a single Pop~III HN event using a horizontal dotted line. The color of the circle symbol corresponds to one of the 30 different MW-like galaxy realizations using {\sc A-SLOTH}. The average total number of minihaloes per MW-like galaxy, <$N_{\rm mini, HNsig}$>, is approximately <$N_{\rm mini, HNsig}$>=583 and <$N_{\rm mini, HNsig}$>=75 for two scenarios considered: one where all Pop~III stars explode as HNe (upper panels), and another where the range for Pop~III HNe is between $M_{\rm \ast, PopIII}=25\msun$ and $M_{\rm \ast, PopIII}=40\msun$ (bottom panels).}
   \label{fig:num_minihalo}
\end{figure}

Figure \ref{fig:num_minihalo} displays the number of progenitor minihaloes containing Pop~III HN signatures in an MW-like galaxy, represented with respect to the halo peak mass (left panels) and stellar mass (right panels) of the MW's satellite analogs. The peak mass refers to the highest halo mass attained by each satellite analog as it moves through the merger tree. The circle symbol in Figure \ref{fig:num_minihalo} represents all 30 realizations of the MW-like galaxy, with a different color for each realization. Especially, Figure \ref{fig:num_minihalo} compares two scenarios: the first one assumes that Pop III star formation with 100\% HN occurrence rate, $f_{\rm HN}=100\%$ (top panels), while the second scenario assumes that Pop~III HNe is triggered within a mass range of $25\msun$ to $40\msun$, giving rise to $f_{\rm HN}=15\%$ (bottom panels). When the results of all 30 runs are considered, the average number of minihaloes with HN signature, denoted as <$N_{\rm mini, HNsig}$>, is found to be <$N_{\rm mini, HNsig}$>$=583$ and <$N_{\rm mini, HNsig}$>$=75$ for the cases with $f_{\rm HN}=100\%$ and $f_{\rm HN}=15\%$, respectively. In the runs where $f_{\rm HN}=1\%$, the average value is notably lower, <$N_{\rm mini, HNsig}$> = 5, while the highest and lowest values are $N_{\rm mini, HNsig}=11$ and $N_{\rm mini, HNsig}=1$, respectively.

The number of progenitor minihaloes with Pop~III HN signature within a satellite galaxy increases as the galaxy mass increases, as illustrated by the linear fit represented by the dotted line in Figure \ref{fig:num_minihalo}. For example, in the scenario with $f_{\rm HN}=100\%$, dwarf galaxies with a stellar mass of $M_{\ast}\gtrsim 10^9\msun$ - similar to the Large Magellanic Cloud (LMC), which has a stellar mass of $M_{\ast}\approx10^{9.1}\msun$ - would contain $N_{\rm mini, HNsig}\approx110$ progenitor minihaloes. However, this number decreases to $N_{\rm mini, HNsig}\approx8$ for galaxies with a mass similar to that of the Sculptor dwarf galaxy, $M_{\ast}\approx10^{6.2}\msun$ (e.g., \citealp{Munoz2018}). In UFD galaxies with stellar masses of $M_{\ast}\lesssim 10^5\msun$, the number of progenitor minihaloes becomes very low, with $N_{\rm mini, HNsig}\lesssim3$. The tendency of more massive galaxies containing a greater number of HN signatures remains consistent even when the frequency of HNe, $f_{\rm HN}$, is decreased. However, the total number of minihaloes is reduced in scenarios where $f_{\rm HN}=15\%$. For instance, for dwarf galaxies with stellar masses of $M_{\ast}\gtrsim10^9\msun$, $M_{\ast}\approx10^6\msun$, and $M_{\ast}\lesssim 10^5\msun$, the corresponding number of minihaloes, $N_{\rm mini, HNsig}$, becomes 11, 2, and less than one, respectively.

\subsection{The likelihood of finding HN signatures}
\begin{figure}
  \centering
  \includegraphics[width=90mm]{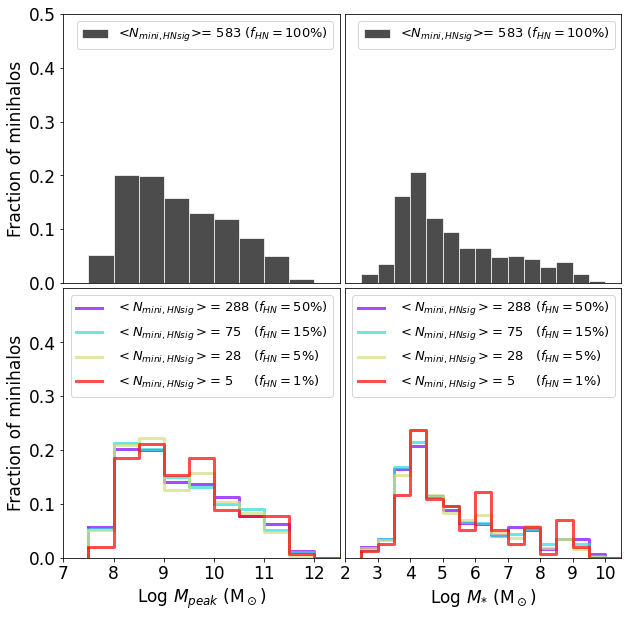}%
   \caption{The fraction of progenitor minihaloes experiencing a Pop~III HN explosion within the satellite galaxies of the MW-like halo, while altering the value of $f_{\rm HN}$, with respect to the halo peak and stellar masses for the satellites at $z=0$. The top panels consider a scenario where all Pop~III stars undergo HNe, while the bottom panels examine cases where the fraction of Pop~III HNe, $f_{\rm HN}$, varies from $1\%$ to $50\%$. Considering the abundances of progenitor minihaloes featuring HN signatures, it is evident that the probability of identifying these minihaloes increases as $f_{\rm HN}$ rises. Meanwhile, when $f_{\rm HN}$ is held constant, the probability of observing progenitor minihaloes with HN signature is higher in smaller haloes compared to larger ones. It means that if an observation is conducted randomly, the chance of discovering a progenitor minihalo is greater in low-mass satellites than in massive ones. This trend is consistent across all scenarios with varying $f_{\rm HN}$. This is due to the fact that {\sc A-SLOTH} is likely to have a higher number of low-mass satellites than massive galaxies.}
   \label{fig:hist_asloth}
\end{figure}

The question then is identifying the satellite galaxy mass at which we are most likely to discover stars with Pop~III HN signatures while searching for a galaxy similar to the MW. To address this question, Figure \ref{fig:hist_asloth} displays the fraction of progenitor minihaloes featuring Pop~III HN signatures within the satellite galaxies of the MW-like halo with respect to the halo peak and stellar mass of satellite analogs at $z=0$. The resulting distribution is based on a series of 30 {\rm A-SLOTH} runs for each scenario, with each run assuming a different value of $f_{\rm HN}$. As a reference, we show the result of the run with $f_{\rm HN}=100\%$ at the upper panels and compare it with other runs where $f_{\rm HN}$ decreases from 50\% to 1\% (bottom panels). Given the prevalence of progenitor minihaloes bearing HN signatures, it becomes apparent that the likelihood of detecting these minihaloes increases as the value of $f_{\rm HN}$ rises.

However, while keeping $f_{\rm HN}$ constant, the likelihood of discovering minihaloes with Pop~III HN signatures is highest in galaxies with low mass, specifically those with halo peak and stellar mass within the range of $M_{\rm peak}=10^{8}-10^{9}\msun$ and $M_{\ast}=10^{3.5}-10^{4.5}\msun$, respectively. For the simulation with $f_{\rm HN}=100\%$, these low-mass galaxies show a fraction of 40\% for containing such minihaloes. As shown in the bottom panels of Figure \ref{fig:hist_asloth}, this trend is also valid for other runs with different $f_{\rm HN}$. For instance, if about 1\% of Pop~III stars undergo HNe, then an MW-like galaxy is predicted to contain a total of 5 minihaloes with the HN characteristics, which are still more likely to be found in small, low-mass satellite dwarfs. It should be, however, emphasized that this trend described above, i.e., the highest probability of finding minihaloes with Pop~III HN signatures in low-mass galaxies, could be simply due to the fact that the {\sc A-SLOTH} model predicts a higher number of low-mass satellites than is observed, as depicted in Figure \ref{fig:Cum_num_sat}. In other words, as demonstrated in Figure \ref{fig:num_minihalo}, a single large galaxy could possess a greater number of minihaloes with Pop III HN signatures than low-mass galaxies, but the total count of low-mass galaxies within the MW volume is higher than that of massive galaxies.

\begin{figure}
  \centering
  \includegraphics[width=90mm]{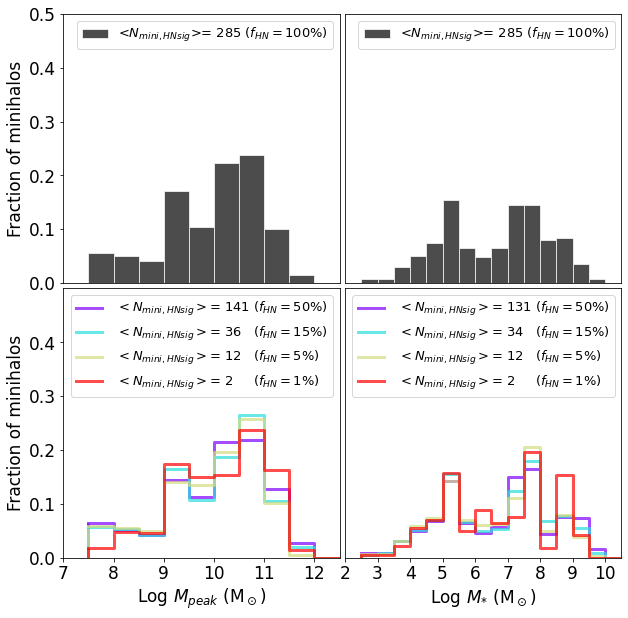}%
   \caption{The same as Figure \ref{fig:hist_asloth}, but it takes into account the actual number of observed galaxies in the MW. As seen in Figure \ref{fig:Cum_num_sat}, {\sc A-SLOTH} predicts more low-mass galaxies than are actually observed. To address this, the fraction of progenitor minihaloes that have undergone HNe is recalculated by adjusting for the difference between the actual observations and the predictions of {\sc A-SLOTH}. This involves decreasing or increasing the number of low and high-mass galaxies by the difference. The result is that the probability of discovering a progenitor minihalo that has experienced HNe with a fixed $f_{\rm HN}$ is highest at halo peak mass and stellar mass of $M_{\rm peak}\approx10^{10.5}\msun$ and $M_{\ast}\approx10^{7.5}\msun$, respectively.}
   \label{fig:hist_obs}
\end{figure}

In order to take into account the discrepancy between the observation and those {\sc A-SLOTH} runs, in Figure \ref{fig:hist_obs}, we adjust the fraction of minihaloes with HN signatures by incorporating this difference. This is achieved by reducing the number of satellite galaxies in the low-mass galaxy regime to approximate the observed number while increasing the number in the high-mass range. Consequently, the distribution of minihaloes with HN signatures exhibits a peak among larger satellite galaxies. The highest fraction values, approximately around 40\%, are found in the region of halo peak masses $M_{\rm peak}\approx10^{10}-10^{11}\msun$ and stellar masses $M_{\ast}\approx10^7-10^8\msun$, respectively. The following significant fraction is approximately 18\%, corresponding to galaxies with halo peak masses $M_{\rm peak}\approx10^9-10^{9.5}\msun$ and stellar masses $M_{\ast}\approx10^5-10^{5.5}\msun$, respectively. Also, we should mention that the total number of minihaloes decreases by a factor of $\sim2$ due to the incorporation of the inconsistency between observations and the {\sc A-SLOTH} simulations. This is because {\sc A-SLOTH} is inclined to predict a higher abundance of low-mass satellite galaxies, which are incomplete in observational data.

In summary, according to the results of the {\sc A-SLOTH} simulation, the most probable dwarf galaxies to contain signatures of HNe associated with Pop~III stars have a stellar mass of $M_{\ast}=10^{3.5}-10^{4.5}\msun$, which is similar to the UFD galaxies like Bootes I, Hercules, CVn II, UMa I, Leo IV, Hydra II, Columba I, Leo V, ComBer, Indus II, UMa II, and Pisces II. 
On the other hand, when accounting for the discrepancy between observations and the {\sc A-SLOTH} runs, galaxies with $M_{\ast}\approx10^7-10^8\msun$, similar to Sagittarius and Fornax, or with $M_{\ast}\approx10^5-10^{5.5}\msun$, corresponding to CVn I, Sextans, Draco, Crater II, are the most probable galaxies to search for to find fossil records of HN signatures of Pop~III stars. 

We should emphasize that the only evidence of an HN signature associated with Pop III stars in the MW satellite galaxies is AS0039 in the Sculptor galaxy. Although {\sc A-SLOTH} simulations suggest that there may be many low-mass dwarfs with HN signatures, the lack of observation in the UFD regime may be due to incomplete observations owing to their dimness. Alternatively, if assuming that the missing satellite problem is solved (e.g., \citealp{Wetzel2015, Read2019, Engler2021}) and we use the results that account for the discrepancy between observations and {\sc A-SLOTH} data, HN signatures are expected to be found in massive satellite galaxies. The discovery of only one HN signature related to Pop~III stars in Sculptor implies that the likelihood of Pop~III stars ending as HN is probably lower than 1\%, according to the {\sc A-SLOTH} simulation, which predicts only two minihaloes with HN signatures assuming $f_{\rm HN}=1\%$. Furthermore, this number of two minihaloes can be considered an upper limit, as our hydrodynamic simulations indicate that only 80\% of Pop~II stars inherit HN signatures when a Pop III HN event occurs. In some cases, the gas may be evaporated or polluted following a Pop III~HN, preventing further star formation or the preservation of HN signatures in subsequent Pop~II stars.

It is important to mention that {\sc A-SLOTH} lacks information regarding the metallicity of stars. As stated in Section 4.2, if a Pop~III star experiences an HN explosion within a minihalo, it is assumed that subsequent Pop II stars born immediately after the explosion might exhibit an HN signature on their surfaces. We plan to delve deeper into this aspect in our forthcoming research. While quantifying the frequency of AS0039-like stars, displaying HN signatures while maintaining low metallicity, among Pop~II stars with HN signatures proves challenging, our hydrodynamic simulations have revealed that generating stars akin to AS0039 requires specific conditions like the rapid infall of pristine gas. Such conditions present difficulties in producing these stars. When considering the amalgamated results of both {\sc A-SLOTH} and our hydrodynamic simulations, our findings suggest that the appropriate fraction of HN in line with observations is likely $f_{\rm HN} \lesssim 1\%$.

\section{Summary and Conclusion}

This study aims to explore the origin of metal-poor stars observed in nearby dwarf galaxies, which could harbor remnants of the earliest generation of stars. Our focus is particularly on AS0039, a star in the Sculptor galaxy, exhibiting distinct features that suggest its association with hypernova (HN) explosions from Pop~III stars. To investigate the conditions under which such stars can form, we have performed six sets of cosmological zoom-in simulations on ultra-faint dwarf (UFD) analogs with a mass of $M_{\rm vir}\approx10^8\msun$ at $z=0$. To explore whether Pop~II stars in our simulations exhibit Pop~III HN signatures, we vary two parameters: the critical metallicity for Pop~II star formation and the fraction of Pop~III stars that undergo HN explosions, defined as $f_{\rm HN}$. The fraction is varied from 100\% to the lower limit where a single HN event occurs during the assembly of the simulated galaxy to cover the full range of possibilities. 

By analyzing the resulting fraction of Pop~II stars exhibiting HN signatures and their metallicity, we have identified the potential environmental conditions for forming metal-poor stars such as AS0039. Furthermore, we also investigate the likelihood of discovering Pop~II remnants that contain HN signatures in nearby dwarf satellite galaxies. To do this, we use a semi-analytic approach with the {\sc A-SLOTH} code that allows us to efficiently explore the parameter space and overcome the limitation of this work, which is confined to a single UFD analog.
\par 

Our main findings are summarized as follows.

\begin{itemize}
    \item According to our simulations, the process of star formation in UFD galaxies is complex, involving the merging of multiple minihaloes. Our results suggest that the main progenitor does not dominate star formation at high redshifts. Instead, we observe that star formation is initiated in minihaloes of comparable mass, which then merge later. Across all simulations, the average in-situ star formation is found to be 19.5\%, and about $\sim$80\% of the stars in the simulated galaxies have grown through mergers.
   
   \item For each simulation run, the number of progenitor minihaloes that contain Pop~II stars exhibiting Pop~III SN signatures, $N_{\rm mini,HNsig}$, varies from zero to 5 depending on the fraction of Pop~III stars that die as HNe. However, the presence of a Pop~III HN event does not guarantee the formation of Pop~II stars with relics of such SN explosions. This can be attributed to the strong feedback from Pop~III HNe, preventing further star formation or the contamination of gas by metals from merged minihaloes prior to the formation of Pop~II stars, leading to the unique HN signatures being diluted and washed out. Moreover, as the fraction of Pop~III HNe decreases, the proportion of Pop~II stars with HN signatures denoted as $f_{\rm PopII,HNsig}$, tends to decrease as well. Specifically, $f_{\rm PopII,HNsig}$ declines from 16\% for {\sc Z55-F100} to 2.3\% for {\sc Z55-Mass}.
   
   \item The metallicity of Pop~II stars within the simulated galaxies ranges of $\rm -5.5<[Fe/H]<-1.5$, while Pop~II stars with HN signatures are typically found in the range from $\rm [Fe/H]\approx-3.3$ to $\rm [Fe/H]\approx-2.4$, with one outlier at $\rm [Fe/H]=-4.17$, similar to the observed {\sc AS0039} star.  We find that the halo, where such ultra metal-poor (UMP) stars are found, undergoes a rapid gas inflow with a peak gas accretion rate of $\dot{M}_{\rm gas}\sim$0.08 $\msun \rm yr^{-1}$, which is significantly higher than the averaged gas inflow rate of $\sim$0.0025 $\msun \rm yr^{-1}$ for other haloes. This suggests that the formation of the UMP star AS0039 in the Sculptor dwarf galaxy could have been facilitated by a halo that experienced a rapid gas inflow. 
   
   \item The metallicity of Pop~II stars formed after Pop~III star formation is highly dependent on the delay time, which refers to the time interval between the Pop~III supernova explosion and the formation of subsequent Pop~II stars. This is due to the fact that, unless there is external enrichment, metallicity usually declines with increased diffusion associated with longer delay times. This could offer another explanation for the formation of Pop~II stars with low metallicities, such as AS0039.
   
   \item The estimated individual metal abundances of Pop~II stars exhibiting HN signatures are in agreement with the best-fit values for AS0039. We confirm that, aside from Pop~III HN events, other progenitors, such as Pop~III stars exploding with a typical SN energy of $E_{\rm SN}=10^{51}$ ergs or Pop~II stars, are unable to produce metal abundances consistent with those found in AS0039. Furthermore, we demonstrate that carbon-enhanced metal-poor stars can be formed via Pop~III stars and that the carbon abundances of Pop~II stars show characteristics that differ from those displaying HN signatures.

   \item Utilizing the semi-analytic {\sc A-SLOTH} code, we find that the average number of minihaloes containing HN signatures in an MW-like galaxy is <$N_{\rm mini}$>$=583$ and <$N_{\rm mini}$>$=75$ for cases where $f_{\rm HN}=100\%$ and $f_{\rm HN}=15\%$, respectively, based on the results of all 30 sets. The number of progenitor minihaloes with HN signatures within a satellite galaxy increases with the galaxy mass. For example, dwarf galaxies with a stellar mass of $M_{\ast}\gtrsim 10^9\msun$, similar to the Large Magellanic Cloud, would contain $N_{\rm mini}\approx110$ progenitor minihaloes if $f_{\rm HN}=100\%$. However, this number decreases to $N_{\rm mini}\lesssim10$ for galaxies with a mass of $M_{\ast}\lesssim10^{6}\msun$.

   \item Our analysis suggests that the likelihood of discovering a progenitor minihalo with HN signatures while maintaining a constant value of $f_{\rm HN}$ is higher in low-mass satellites ($M_{\rm peak}=10^{8}-10^{9}\msun$ and $M_{\ast}=10^{3.5}-10^{4.5}\msun$) compared to more massive ones. However, we also find that the semi-analytic {\sc A-SLOTH} code tends to over-predict the number of low-mass satellites compared to what is observed in the local Universe. This discrepancy could be due to observational incompleteness caused by the faintness of low-mass galaxies. Taking this into account, we find that the most likely dwarf galaxies to contain HN signatures are shifted towards more massive satellite galaxies, with a probability of 40\% in the range of $M_{\rm peak}\approx10^{10}-10^{11}\msun$ and $M_{\ast}\approx10^7-10^8\msun$. The subsequent highest probability of discovering minihaloes with HN signatures is associated with galaxies in the range of $M_{\rm peak}\approx10^9-10^{9.5}\msun$ and $M_{\ast}\approx10^5-10^{5.5}\msun$.
   
\end{itemize}
It is worth noting that the simulated galaxy in this study has less mass, approximately two orders of magnitude lower than the Sculptor galaxy, where AS0039 was discovered. The Sculptor galaxy is classified as a dwarf spheroidal (dSph) galaxy and has a mass of $M_{\ast}\approx1.8\times10^6\msun$. Nevertheless, the results of this study indicate that stars like AS0039 could have formed during the early stages of the evolution of the Sculptor galaxy. This is supported by the fact that the formation of Pop~III stars occurs early in the Universe with a cosmic time of $t_{\rm H}\lesssim0.75$ $\rm Gyr$, and the remnants of Pop~III stars are only found in Pop~II stars that formed shortly after. This suggests that early star formation is a key factor in producing stars with characteristics similar to AS0039, regardless of whether the galaxy is a dSph or not. However, the analysis using {\sc A-SLOTH} indicates that the probability of discovering HN signatures in dwarf galaxies is dependent on the number of satellite galaxies with a specific mass. Given the current limitations in observations, it is more probable to find stars with relics from the first generation of stars in more massive satellite galaxies.

Considering the current limitations in observations, the discovery of only one star, AS0039, among the observed satellite galaxies suggests that the fraction of Pop~III stars that undergo HN explosions is extremely low. According to the {\sc A-SLOTH} analysis, assuming that only 1\% of Pop~III stars die as HNe, only two minihaloes are predicted to contain Pop~II stars with Pop~III HN signatures, implying the fraction of Pop~III stars exploding as HNe likely to be less than 1\%. According to \citet{Yoon2012}, Pop~III stars with masses between $13\msun$ and $83\msun$ could experience HN explosions depending on their degree of rotation. Certainly, the fraction of Pop~III stars that end in HNe is determined by the assumed initial mass function (IMF). Our adopted IMF predicts an $f_{\rm HN}\approx54\%$ over the possible progenitor masses for HN under chemically homogeneous evolution. Conversely, our results, when combined with such nucleosynthesis findings, could provide insights into the rotational degree of the first generation of stars.

Despite the JWST being a powerful tool for studying early objects formed in the early Universe, it is anticipated that directly observing individual first-generation stars will be extremely challenging (e.g., \citealp{Schauer2020, Woods2021, Katz2023, Larkin2023}). Consequently, the role of stellar and galactic archaeology in inferring the nature of the first stars will be crucial, providing a complementary understanding of far-field cosmology. Detailed chemical abundances in metal-poor stars within local dwarf galaxies can provide insight into questions such as the end of life of the first stars, their mass and spin, and their birth environment. Moreover, the near-field cosmology approach will be bolstered by future observations using advanced telescopes like the Giant Magellan Telescope (GMT), the Thirty Meter Telescope (TMT), and the European Extremely Large Telescope (E-ELT), which promise to offer improved spectroscopic sensitivity to uncover the chemical signature left behind by early cosmic history.


\section*{acknowledgements}
We express our gratitude to the anonymous referee for their constructive and insightful feedback, which has significantly enhanced the clarity of our paper. 
We thank Hartwig Tilman for the valuable discussions and for graciously sharing the {\sc A-SLOTH} code with us. We are grateful to Volker Springel, Joop Schaye, and Claudio Dalla
Vecchia for permission to use their versions of \textsc{gadget}.
T.~L. and M.~J. are supported by the National Research Foundation (NRF) grants No. 2021R1A2C109491713 and No. 2022M3K3A1093827, funded by the Korean government (MSIT).

\section*{DATA AVAILABILITY}
The simulation data and results of this paper may be available upon
request.

\bibliography{myrefs2}{}
\bibliographystyle{mnras} 

\appendix
\section{EVOLUTION OF THE TARGET HALO}

\begin{figure}
    \centering
    \includegraphics[width=90mm]{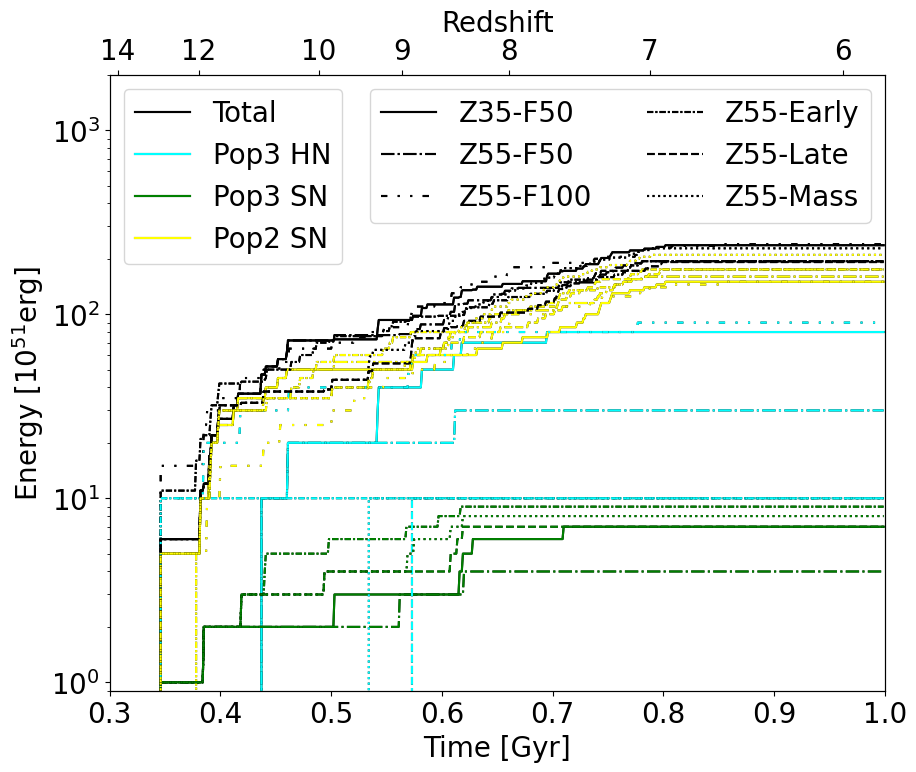}
    \caption{The accumulation of feedback energy from both SNe and HNe exerted on the target halo is depicted for the six simulation sets. The feedback energy arising from Pop~III SNe is represented by the green line, Pop~III HNe by the cyan line, Pop~II SNe by the yellow line, and the cumulative total feedback energy by the black line. Each individual simulation set is differentiated by a different line style. Interestingly, despite the higher energy release of a Pop~III HN ($E_{\rm SN}=10^{52}$ erg) compared to a Pop~III SN ($E_{\rm SN}=10^{51}$ erg) and Pop~II SNe ($E_{\rm SN}=5\times10^{51}$ erg), the overall energy contribution is more likely to be dominated by Pop~II SNe due to the infrequent occurrence of Pop~III HNe. Across all six simulation sets, the target halo ultimately converges to a comparable baryon fraction, as demonstrated in the left panel of Figure \ref{fig:Mass_SFH}.}
    \label{fig:SN_energy}
\end{figure}

\begin{figure}
  \centering
  \includegraphics[width=90mm]{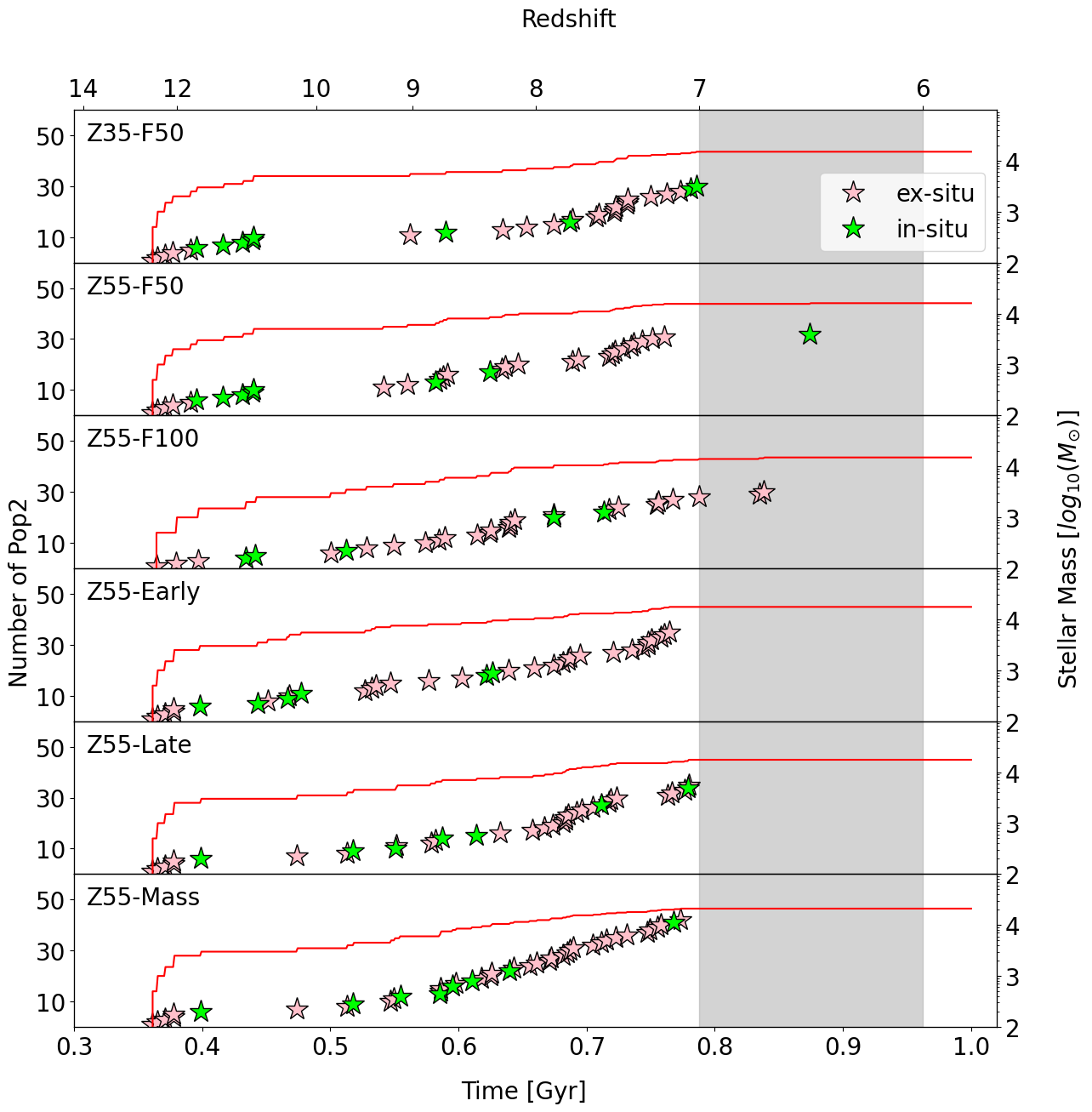}%
   \caption{The formation history of Pop~II stars within the target halo. The number of stars originating in-situ within the main progenitor halo are denoted by green star symbols, while ex-situ stars from other progenitor haloes are represented by pink star symbols. The right y-axis is dedicated to showing the stellar mass, represented by the continuous red line, identical to the red lines found in the left panel of Figure \ref{fig:Mass_SFH}.}
   \label{fig:insitu_pop2}
\end{figure}

In Figure \ref{fig:SN_energy}, we have analyzed the amount of feedback energy from both SNe and HNe on the target halo across cosmic time. Note that the energy associated with a single Pop~III SN is $E_{\rm SN} = 10^{51}$ erg, while for a Pop~III HN, it is $E_{\rm SN} = 10^{52}$ erg. For a single Pop~II star particle, represented by a stellar mass of $500 M_{\odot}$, the SN energy is equivalent to $E_{\rm SN} = 5 \times 10^{51}$ erg. Figure \ref{fig:insitu_pop2} illustrates the formation history of Pop~II stars within the target halo by tracing their origins. In-situ stars originating from the primary progenitor halo are indicated by green symbols, while those emerging from other progenitor haloes are denoted by pink symbols.

\section{MERGER HISTORY OF THE TARGET HALO}

\begin{figure}
    \centering
    \includegraphics[width=80mm]{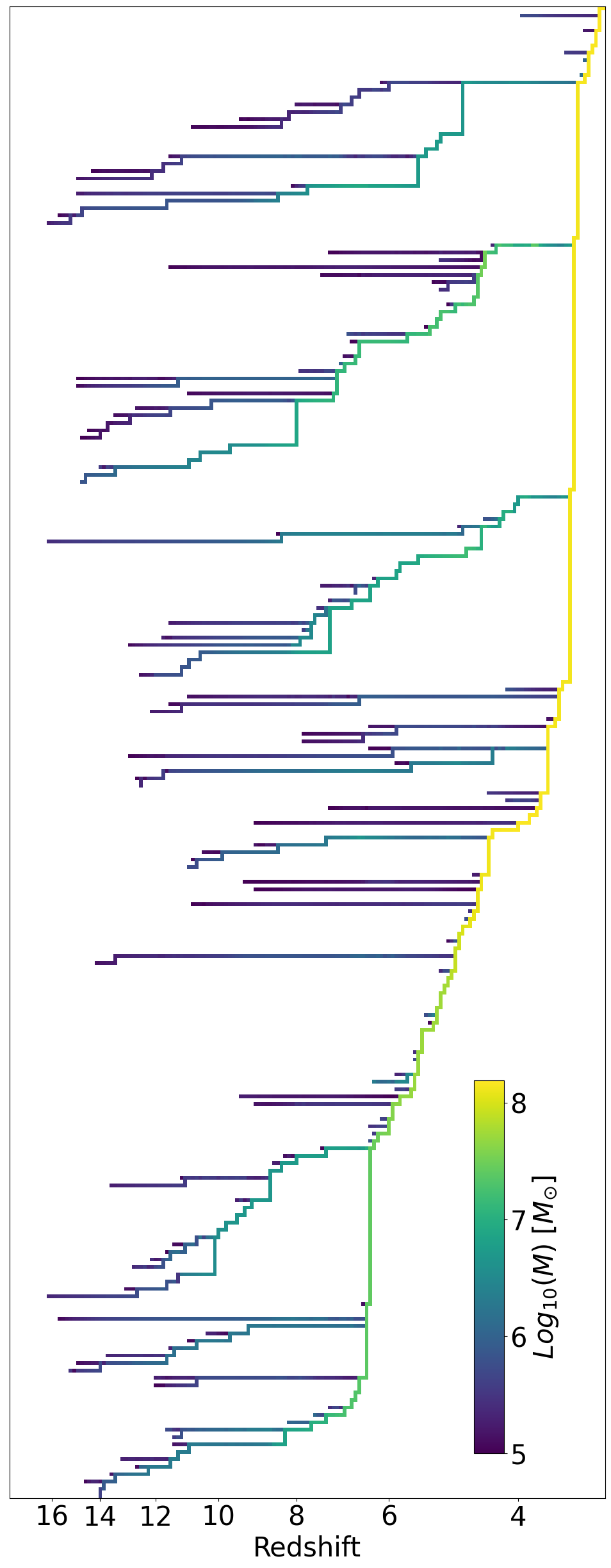}
    \caption{The merger tree of the target halo within the {\sc Z55-F50} simulation at $z \approx 3$. Each line on the diagram represents a progenitor halo with a dark matter mass exceeding $M_{\rm vir}\approx10^{5} M_{\odot}$ and is color-coded in accordance with its mass. Specifically, haloes with $M_{\rm vir}>10^6\msun$ are represented in green, transitioning to yellow as the mass increases to around $M_{\rm vir}\approx10^8\msun$. Despite the target halo being formed from a composite of multiple progenitor haloes, the number of progenitor haloes containing Pop~II stars with Pop~III signatures is estimated to be approximately $N_{\rm mini, PopIII}\approx10$. These stars are predominantly formed within the mass range of haloes $10^6
    \msun \lesssim M_{\rm vir}\lesssim 4\times10^6\msun$, consistent with the findings depicted in Figure \ref{fig:n_H_minihalo}.
    }
    \label{fig:mtree}
\end{figure}








\bsp	
\label{lastpage}
\end{document}